\documentclass[preprint,showpacs,showkeys,aps,prd,amsfonts,eqsecnum,nofootinbib,floatfix]{revtex4}
\usepackage{graphicx,epsfig}
\usepackage[dvips]{color}
\usepackage{amsmath, amssymb, graphics}

\newcommand{\gen}[3]{\ensuremath{H^{#1}\to \chi^{#1}_#2 \chi^0_#3}}

\def\gsim{\lower0.5ex\hbox{$\:\buildrel >\over\sim\:$}}
\def\lsim{\lower0.5ex\hbox{$\:\buildrel <\over\sim\:$}}
\begin{document}

\preprint{CUMQ/HEP 145}
%
%
\title{\Large  CP Asymmetry in Charged Higgs Decays to Chargino-Neutralino}
\author{Mariana Frank}\email[]{mfrank@alcor.concordia.ca}
\author{Ismail Turan}\email[]{ituran@physics.concordia.ca}
\affiliation{Department of Physics, Concordia University, 7141
Sherbrooke Street West, Montreal, Quebec, CANADA H4B 1R6}
\date{\today}

\begin{abstract}
We analyze the charge-parity (CP) asymmetry in the charged Higgs boson decays to chargino-neutralino pairs, 
\gen{\pm}{i}{j},$~i=1,2,\,j=1,\hdots 4$. We show first that these modes have a large branching ratio 
for $m_{H^{\pm}} \gtrsim 600$ GeV. We use Cutkosky rules to obtain the analytical formulas needed for the evaluation of the asymmetry under consideration. We then calculate the CP asymmetry in chargino-neutralino decays by including supersymmetric mass bounds, as well as constraints from $b \to s \gamma$, $(g-2)_{\mu}$, $\Delta\rho$ and electric dipole moments. Finally, we discuss observability of the asymmetry at the LHC by calculating the number of required charged Higgs events to observe the asymmetry for each decay channel. We show that the inclusion of constraints considerably reduces the projected CP asymmetry, and that the optimal channel for observing the  asymmetry is \gen{\pm}{1}{2}.
\end{abstract}
\maketitle
\section{Introduction}\label{sec:intro}
The Large Hadron Collider (LHC) at CERN is expected to unravel the mysteries of electroweak symmetry breaking  as well as provide signals of physics beyond the Standard Model (SM). One would hope that in the first instance, the Higgs boson would be found. In the SM, there is only one physical (neutral) Higgs boson. Should a charged scalar particle be found, one would have to unravel the underlying physics, as most frameworks beyond SM predict charged Higgs bosons. 
The simplest  and the most popular of such models are the two Higgs doublet model(s) and supersymmetric models. One could distinguish among the ``new Higgs'' by testing their charge-parity (CP) properties. In the (SM) the CP breaking occurs through only one weak phase in Cabibbo-Kobayashi-Maskawa (CKM) matrix. The observed direct CP violations in K \cite{Caso:1998tx} and B \cite{Aubert:2001fg} decays can be 
accommodated via the CKM matrix in the SM.  Models beyond the SM contain several new sources  of CP violation, some of which  could affect Higgs boson decays. The minimal supersymmetric standard model (MSSM) contains many parameters which can in principle be complex, even after making all 
allowed rotations to get rid of unphysical phases. 

Such new sources of flavor and CP violation give rise to enhanced CP violation, some of which could provide distinguishing signs for MSSM at present and future colliders. In a previous work \cite{Frank:2007ca}, we discussed the CP asymmetry in charged Higgs decays $H^-\to\bar{u}_i d_j$ in
the framework of the MSSM, where $\bar{u}_i d_j = \bar{c} b, \bar{c} s, \bar{t} b, \bar{t} s$. We 
showed that, although  the channels $\bar{t} s$ and 
$\bar{c} s$ have sizable CP asymmetry with respect to the channels $\bar{t} b$ and $\bar{c} b$, 
this result has to be taken with caution since the formers have very small branching ratios ($Br$'s) 
which makes them harder to observe. We calculated $(A_{CP}^2 \times Br)^{-1}$ \cite{Eilam:1991yv} for each decay mode, which is proportional to the  number of 
required charged Higgs bosons produced at colliders\footnote{$(A_{CP}^2 \times Br)^{-1}$ is closely related to the total number of events $N$ required to establish a measurable CP violation for a particular mode. The exact formula is $N=s^2(A_{CP}^2 \times Br\,\epsilon)^{-1}$. Here $s$ is the standard deviation and $\epsilon$ is the detection efficiency.} to observe an asymmetry for a given channel \cite{Eilam:1991yv,Christova:2006fb,Atwood:2000tu}. Based on this analysis, we concluded that  $H^- \to \bar{c} b$ and $H^- \to \bar{t} b$ are the 
the optimal channels which could reveal a 
measurable CP asymmetry at the order of $10 - 15\%$. 

In this study, we calculate the CP asymmetry in charged Higgs decays  in the purely supersymmetric mode \gen{\pm}{i}{j}, $i=1,2,\,j=1,\hdots 4$, compare the  size of the CP asymmetry for each $i$ and $j$, and evaluate the number of charged Higgs needed to observe each asymmetry. We show that the asymmetry can be very large if we include only constraints requiring  supersymmetric particles to have masses consistent with experimental bounds. However, if we also include low energy phenomenology constraints coming from $b \to s \gamma$, $(g-2)_{\mu}$, $\Delta\rho$ and electric dipole moments, the predicted asymmetry and the allowed parameter space are considerably reduced.
 
The outline of the paper is as follows. In the next section, we briefly review the CP structure of the MSSM. The 
processes and the method of calculation are outlined in Section \ref{decays} and the numerical analysis of the decays under consideration are presented in the Section \ref{numerical}. Finally, we conclude 
in Section \ref{conc}, and give the relevant analytical expressions in the Appendix.

\section{The CP structure in the chargino and neutralino sectors of the MSSM}
In addition to the CP violation induced by the CKM matrix  in the quark charged current, CP phases  can be introduced explicitly in the MSSM, by complex Yukawa couplings of Higgs bosons to quarks and squarks. New supersymmetric-only sources of CP violation in MSSM are: the soft supersymmetry breaking gaugino masses $M_\alpha, \,\alpha=1,2,3$; the bilinear Higgsino coupling parameter $\mu$; and the soft supersymmetry breaking trilinear scalar coupling of the Higgs bosons with scalar fermions $f$, $A_f, \,f=u,d,c,s,t,b,e,\mu,\tau$. In principle, each of these parameters can have independent  CP-phases, making a general analysis of CP violation or asymmetry very complicated. We make several simplifying assumptions. First, we assume that at the unification scale the gaugino masses have a common phase and the 
trilinear couplings are  all equal and have another common phase. 
In order to avoid known problems with the Electric
Dipole Moments (EDMs) \cite{Chang:1998uc}, one could deviate from exact universality and consider $A_f$ 
to be diagonal in flavour space with vanishing first and second generation
couplings. 
This leaves independent phases in $\mu,M_\alpha$ and $A_f$. 
However, the symmetries of the
MSSM can be used to re-phase one of the Higgs doublet fields and the
gaugino fields such that $M_\alpha$ are real 
\cite{Pilaftsis:1999qt,Dugan:1984qf}. In addition, the CP-violating phases associated with the sfermions of the first and second generations are severely constrained by bounds on the 
EDMs of the electron, neutron and muon \cite{Fischler:1992ha} and this limits $\arg(\mu)$, which cannot exceed values of the order of $10^{-2}-10^{-3}$.

There have been several proposals \cite{Nath:1991dn}--\cite{Ibrahim:1998je} to evade these 
constraints without suppressing the CP-violating phases. One possibility is to 
arrange for partial cancellations among various contributions to the 
EDMs \cite{Ibrahim:1998je}.
Another option is to make the first two generations 
of scalar fermions rather heavy, of order $3$ TeV, so that the one-loop 
EDM constraints are automatically avoided. One could invoke the
 effective SUSY models \cite{Dimopoulos:1995mi} where decoupling of the first 
and second generation sfermions are used to solve the SUSY Flavour Changing 
Neutral Current (FCNC) and CP 
problems without spoiling the naturalness condition. We adopt the latter 
version of a CP-violating MSSM for our analysis, along with $A_f=0$ for 
the first two generation sfermions.  We also simply neglect $\arg(\mu)$ and 
 consider $\mu$ real.  Thus, we assume that the only  non-zero phases emerge from the trilinear 
couplings $A_{t,b,\tau}$ and  we assume a common phase\footnote{ 
The CP-violating phases $\arg(\mu)$ and $\arg(A_{t,b})$ could
in principle be measured directly in the production cross sections and
decay widths of (s)particles in high energy colliders \cite{Pilaftsis:1999qt},
\cite{Choi:1999aj}- \cite{Carena:2001fw} or indirectly via their radiative effects
on the Higgs sector \cite{Pilaftsis:1999qt,Choi:2001pg}.} $\arg(A_t) = \arg(A_b)=\arg(A_{\tau}) \equiv \arg(A)$ unless otherwise stated. 

The superpotential $\mathcal{W}$ of the MSSM Lagrangian  and the relevant 
part of the soft breaking Lagrangian 
$\mathcal{L}^{\text{squark}}_{\text{soft}}$ are respectively 
\begin{eqnarray}
     \label{eq:W} 
\!\!\!{\mathcal{W}} &=& \mu H^1 H^2 + Y_l^{ij} H^1
{L}^i {e}_R^j + Y_d^{ij} H^1 {Q}^i {d}_R^j
+ Y_u^{ij} H^2 {Q}^i {u}_R^j\\
 \!\!\!\!\!\!\!\mathcal{L}^{\text{squark}}_{\text{soft}}\!\!\!\! &=&\!\!\!
-\tilde Q^{i\dagger} (M_{\tilde Q}^2)_{ij} \tilde Q^j
-\tilde u^{i\dagger} (M_{\tilde U}^2)_{ij} \tilde u^j
-\tilde d^{i\dagger} (M_{\tilde D}^2)_{ij} \tilde d^j 
+ Y_u^i A_u^{ij} \tilde Q_i H^2 \tilde u_j
+ Y_d^i A_d^{ij} \tilde Q_i H^1 \tilde d_j,
\label{eq:superpot}
\end{eqnarray}
where $H^1$, and $H^2$ are the Higgs doublets with vacuum expectation values 
$v_1$ and $v_2$ respectively, $ Q$ is the $SU(2)$ scalar 
doublet, $ u$, $ d$ are 
the up- and down-quark $SU(2)$ singlets, respectively,  $\tilde Q, \tilde u, \tilde d$
represent scalar quarks,  $Y_{u,d}$ are the
Yukawa couplings and $i,j$ are generation indices. Here $A^{ij}$ represent the 
trilinear scalar couplings.  

The CP violation effects enter the Higgs decays to chargino and neutralino at one loop only through loops containing scalar quarks. For simplicity, and to avoid introducing new parameters, we neglect flavor mixing in the scalar quark mass matrix. The scalar quark mass is 
taken as:
\begin{equation}
\label{eq:squarkmass}
\!\!\!\!\!\!\!\!{\cal M}^2_{\tilde {t}\{\tilde b\} }=
\left( \begin{array}{cc}
M_{{\tilde L} t\{b\}}^2 & m_{t\{b\}} {\cal A}_{t\{b\}}
 \\
m_{t\{b\}} {\cal A}_{t\{b\}}^{\ast} & M_{{\tilde R} t\{b\}}^2 
\end{array} \right)
\end{equation}
with
\begin{eqnarray}
\label{eq:squarkparam}
M_{{\tilde L}q}^2 &=&
      M_{\tilde Q,q}^2 + m_q^2 + \cos2\beta (T_q - Q_q s_W^2) M_Z^2\,,\;
\nonumber \\
{\cal A}_{t} &=& A_{t} - \mu\cot\beta\,,\;\;\;
{\cal A}_{b} = A_{b} - \mu\tan\beta\,. 
\end{eqnarray}
The mass squared matrix for the scalar $\tau$ lepton is similar to the one for b-squark, with the appropriate substitutions. Here we assume $\mu$ real and a common phase for $A_{t}$ and $A_{b}$. Note that we neglect the effects of 
$A_{c,s}$ since they are multiplied by the charm or strange quark mass. Here $\tan\beta$ is the ratio of the vacuum expectation values of the two neutral Higgs bosons.

We summarize the basic properties of the charginos and neutralinos entering in the calculation of the CP asymmetries, as well as give the basic couplings between them and the scalar quarks entering the loop calculations. For more information, see \cite {Frank:2006ku}. 

The charginos $\chi_i^+\; (i=1,2)$ are four component Dirac
fermions. The mass eigenstates are mixtures of the winos
$\tilde{W}^\pm$ and the charged higgsinos $\tilde{H}^-_1$,
$\tilde{H}^+_2$. 
The chargino masses are eigenvalues of the
diagonalized mass matrix,
 {\bf X}, given by
\begin{equation}
{\bf X} = \left( \begin{array}{cc} M_2 & \sqrt2\, M_W \sin \beta \\[1ex] \sqrt2\, M_W\, 
          \cos \beta & \mu  \end{array}\right)~,
\end{equation}
where $M_2$ is the soft SUSY-breaking $SU(2)_L$ gaugino mass parameter and $\mu$ is the Higgsino mass parameter. 
The relevant Lagrangian terms for the
     quark-down squark-chargino 
interaction are given by
\begin{eqnarray}
\mathcal{L}_{u\tilde{d}\chi^{+}}\!\!\!&=\!\!\!&\sum_{\sigma=1}^{2}\,
\sum_{i,j=1}^{3}\left\{ \bar{u}
_ {i}\,[V_{\sigma 2}^{*}\,(Y_{u}^{\text{diag}}\,K_{CKM})_{ij}]
\,P_L\,\chi
_{\sigma}^{+}\,(\Gamma_D)^{ja}\,\tilde{d}_{a}-\bar{u}_{i}\,[g\,U_{\sigma
1}\,(K_{CKM})_{ij}]\, P_R\,
\chi_{\sigma}^{+}\,(\Gamma_D)^{ja} \,\tilde{d}_a\right.   \nonumber \\
& &  \left. +\,\bar{u}_{i}\,[U_{\sigma 2}\,(K_{CKM}\,Y_{d}^{%
\text{diag}})_{ij}]
\,P_R\,\chi_{\sigma}^{+}\,(\Gamma_D)^{(j+3)a}\,\tilde{d}_a
\right\} +\text{%
H.c.}\,,
\end{eqnarray}
where the index $\sigma$ refers to chargino mass eigenstates.
$Y_{u,d}^{\text{diag}}$ are the
diagonal up- and down-quark Yukawa couplings, and $V$, $U$ the 
chargino rotation matrices defined by $U^{*}{\bf X}\,V^{-1}=\mathrm{diag}%
(m_{\chi_{1}^{+}},m_{\chi_{2}^{+}})$. 

Neutralinos $\chi_i^0\; (i=1,2,3,4)$ are four-component
Majorana fermions. The mass eigenstates are mixtures of the
photino,~$\tilde{\gamma}$, the zino,~$\tilde Z$, and the neutral higgsinos,
$\tilde{H}^0_1$ and $\tilde{H}^0_2$. Their mass matrix {\bf Y} is given by
\begin{equation}
\renewcommand{\arraystretch}{1.2}
\!\!\!\!{\bf Y} = \left( \begin{array}{cccc} M_1 & 0 & -M_Z\sin \theta_W\cos \beta & M_Z\sin \theta_W\sin \beta \\ 0 &
          M_2 & M_Z\cos \theta_W\cos \beta & -M_Z\cos \theta_W\sin \beta \\
          -M_Z\sin \theta_W\cos \beta & M_Z\cos \theta_W\cos \beta & 0 & -\mu \\ M_Z\sin \theta_W\sin \beta &
          -M_Z\cos \theta_W\sin \beta& -\mu & 0\end{array}\right)\,.
\label{Y}
\end{equation}
The relevant Lagrangian terms for the quark-up squark
neutralino interaction are
\begin{eqnarray}
\mathcal{L}_{u\tilde{u}\chi^{0}}&=&\sum_{n=1}^{4}\sum_{i=1}^{3}\left\{
\bar{u}
_{i}\,N_{n1}^{*}\,\frac{4}{3}\frac{g}{\sqrt{2}}\tan \theta _{W} \,P_L\,%
\chi_{n}^{0}\,(\Gamma_U)^{(i+3)a}\,\tilde{u}_a-\bar{u}_{i}\,N_{n4}^{*}\,Y_{u}^{\text{%
diag}}\,P_L\,\chi_{n}^{0}\,(\Gamma_U)^{ia}\,\tilde{u}_a
\right.   \nonumber \\
&-& \left.\bar{u}_{i}\,\frac{g}{\sqrt{2}}\left( N_{n2}+%
\frac{1}{3}N_{n1}\tan \theta _{W}\right)\,P_R
\,\chi_{n}^{0}\,(\Gamma_U)^{ia}\,\tilde{u}_a%
-\bar{u}_{i}\,N_{n4}\,Y_{u}^{\text{diag}}\,P_R\,\chi_{n}^{0}\,(\Gamma_U)^{(i+3)a}\,%
\tilde{u}_a\right\} \,,\nonumber \\
\end{eqnarray}
where $N$ is the rotation matrix which diagonalizes the
neutralino mass matrix $M_{\chi^0}$, $N^{*}{\bf Y}\,N^{-1}=\mathrm{diag}(m_{%
\chi_{1}^{0}},\,m_{\chi_2^0}, \,m_{\chi_3^0}, \,m_{\chi_4^0})$.

 In all of the above interactions, CP violation arise from $\arg(A_{t,b})$. Though we include squark contributions only for simplification, sleptons contribute too in an analogous manner. We check the sensitivity of the CP asymmetry to  $\arg(A_{\tau})$ as well.

\section{Charged Higgs decays \gen{\pm}{i}{j} and CP asymmetry}
\label{decays}
\begin{figure}[b]
        \centerline{\epsfxsize 6.5in {\epsfbox{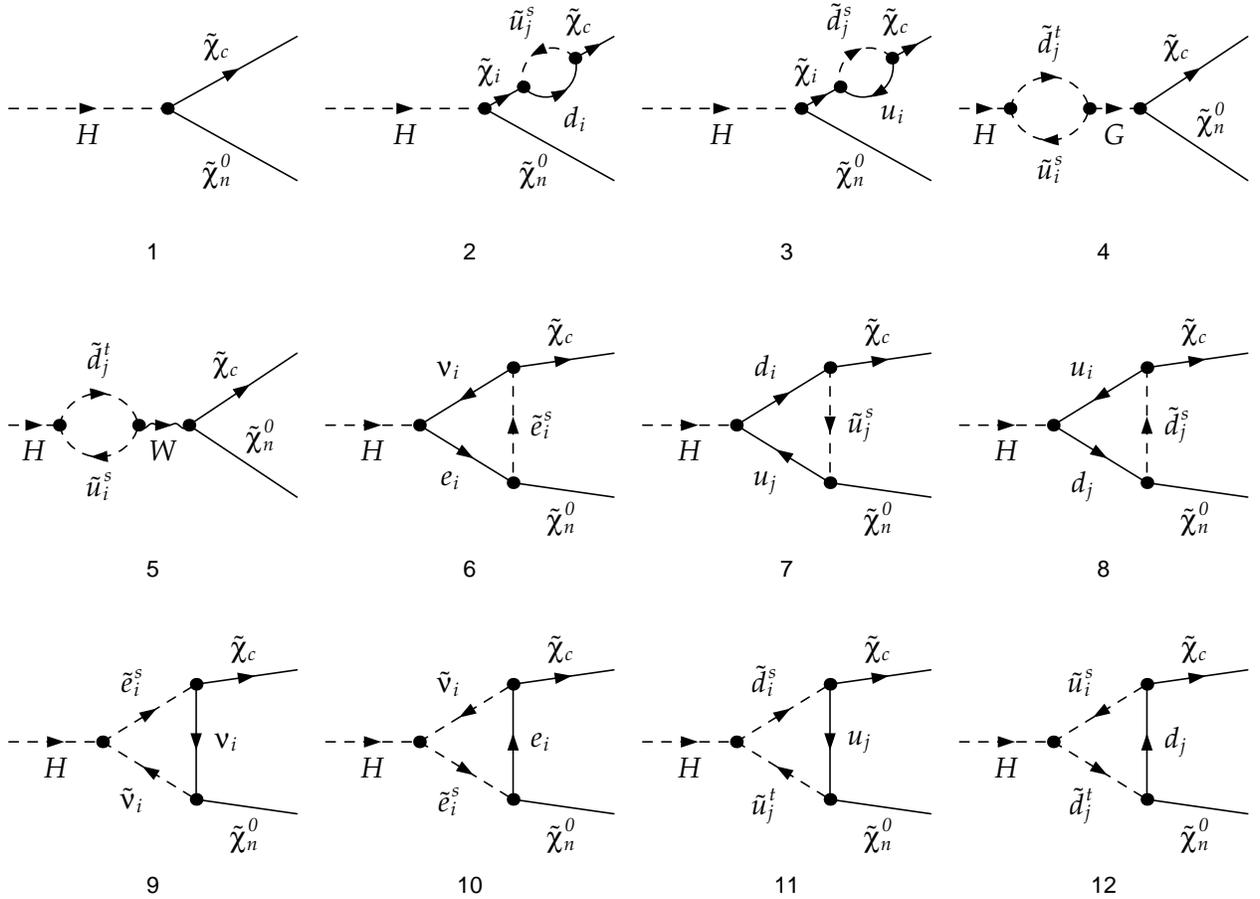}}}
\caption{The tree and relevant one-loop diagrams contributing CP asymmetry in t' Hooft-Feynman gauge for the decays \gen{-}{c}{n}.}
\label{fig:relevant}
\end{figure}
The relevant Feynman diagrams for calculating the CP asymmetry in charged Higgs decays to chargino-neutralino pairs are shown in Fig.~\ref{fig:relevant}. There are indeed more than 120 Feynman diagrams contributing at one-loop level but based on our choice of the CP phases the list goes down to eleven diagrams given in  Fig.~\ref{fig:relevant}.

The CP asymmetry is defined as:
\begin{equation}
\displaystyle A_{CP} =\frac{\Gamma(\gen{-}{i}{j})-\Gamma(\gen{+}{i}{j})}{\Gamma(\gen{-}{i}{j})+\Gamma(\gen{+}{i}{j})},
\label{ACP} 
\end{equation}
For simplicity, in the denominator of Eq.~(\ref{ACP}), we approximate the decays width of the charged Higgs by their tree-level contribution since only the real part of the decay width is needed, and including one-loop contributions is not essential.
\begin{figure}[htb]
\hspace*{-0.3cm} 
\centerline{\epsfxsize 6.2in {\epsfbox{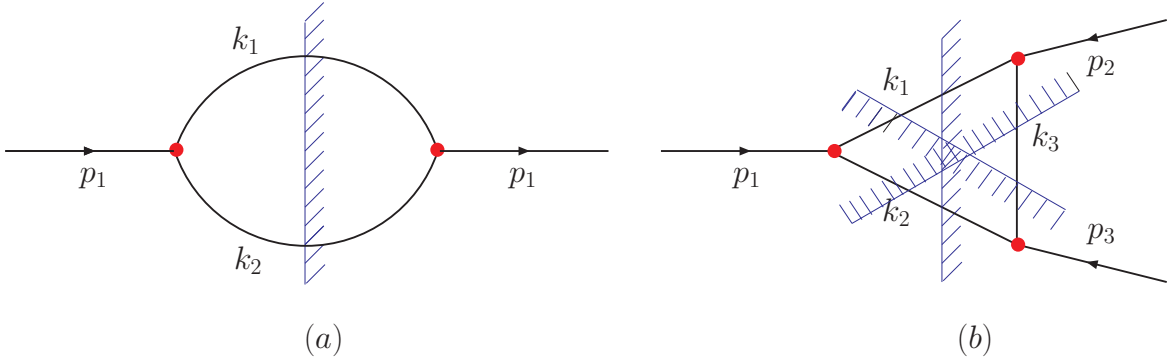}}}
\vspace*{-0.1in}
\caption{The unitarity cuts for self energy $(a)$ and vertex type $(b)$ generic diagrams.}
\label{fig:cuts}
\end{figure}

The CP-odd observable $A_{CP}$ requires a nontrivial phase from Feynman diagrams (called absorptive or strong phase),  in addition to the weak phase from $\arg(A_{t,b,\tau})$ mentioned in the previous section. This way, the imaginary part of the amplitude is non-zero, ${\mathcal Im}(\rm Amplitude)\ne 0$. One way of introducing such a phase is through one-loop Feynman diagrams, where some of the intermediate particles go on-shell. As tree-level contributions do not generate a CP asymmetry by themselves, we estimate the CP asymmetry from the interference of tree and one-loop diagrams, which dominate over corresponding contributions from loop-loop terms. We show in Fig.~\ref{fig:relevant} the tree level and the one-loop graphs which only contribute to the asymmetry ({\it i.e.}, the ones with scalar fermions in the loop). 

One convenient way of calculating CP asymmetry is to extract the absorptive part of the loop diagrams by applying the Cutkosky rules \cite{Cutkosky:1960sp}. For self-energy type diagrams there is one possible cut, but for the vertex diagrams there are three types of cuts depending on the masses of the external particles. So, in Fig.~\ref{fig:cuts}, we show possible cuts where the {\it internal-cut} states represent all possibilities  in each case. For the details of the method of calculation, see Ref.~\cite{Frank:2007ca}. 

For self-energy diagrams the vertical one can be: a cut through scalar top - down quark, scalar bottom - up quark, or scalar top - scalar bottom loop. They contribute if 
 $m_{\chi_c^-}\geqslant m_{\tilde{t}}+m_{d_j}$,  $m_{\chi_c^-}\geqslant m_{\tilde{b}}+m_{u_j}$, or 
$m_{H^-}\geqslant m_{\tilde{t}}+m_{\tilde{b}}$, respectively. Here $\chi_c^-$ represents the final state chargino with $c=1,2$. 

For the vertex diagrams, as seen from Fig.~\ref{fig:cuts}, there are three possible cuts: one vertical, two horizontal ones. The vertical cuts through $\tau-\nu_\tau$, $b-t$, and $t-b$ always contribute if a non-zero phase for $A_\tau$, $A_t$, and $A_b$ is assumed, respectively, and for charged Higgs masses $m_{H^\pm}>m_t$. The other vertical cuts through $\tilde{\tau}-\tilde{\nu}_\tau$ and  $\tilde{t}-\tilde{b}$ depend on the chosen Higgs mass as well as the phases of $A_{t,b,\tau}$. The situation is quite similar for the other two horizontal cuts, but this time the mass of the chargino or neutralino has to be greater than the sum of the masses of the internal particles where we cut through. Some analytical formulas are given in Appendix for the horizontal and vertical cuts in a generic way, rather than giving lengthy analytical results for the imaginary parts.

We generate the amplitudes using $\tt FeynArts$ \cite{Hahn:2000jm} and proceed with our own code for the rest of the calculation. In the numerical calculation we checked our results for the imaginary parts of the two-point and three point diagrams with the package program {\tt LoopTools} \cite{Hahn:2000jm} and found close agreement. Also, for the numerical diagonalizations of chargino, neutralino and scalar quark mass matrices, we cross checked our routines with the Hahn's routines {\tt Diag} \cite{Hahn:2006hr}.

\section{Numerical Analysis}
\label{numerical}
In this section we present a comparative numerical analysis of the CP asymmetry in charged Higgs decays into all possible pairs of chargino and neutralino states. Our aim is two-fold:  the first is to predict the CP asymmetry and indicate the optimal channel as well as estimate the number of charged Higgs needed to probe it. The second is to analyze the effect of various experimental constraints such as $b \to s \gamma$, $(g-2)_{\mu}$, $\Delta\rho$ and EDMs and determine their significance. 

\begin{figure}[htb]
\begin{center}$
\hspace*{-1.3cm}
	\begin{array}{c@{\hspace{-1cm}}c}
	\includegraphics[width=3.75in]{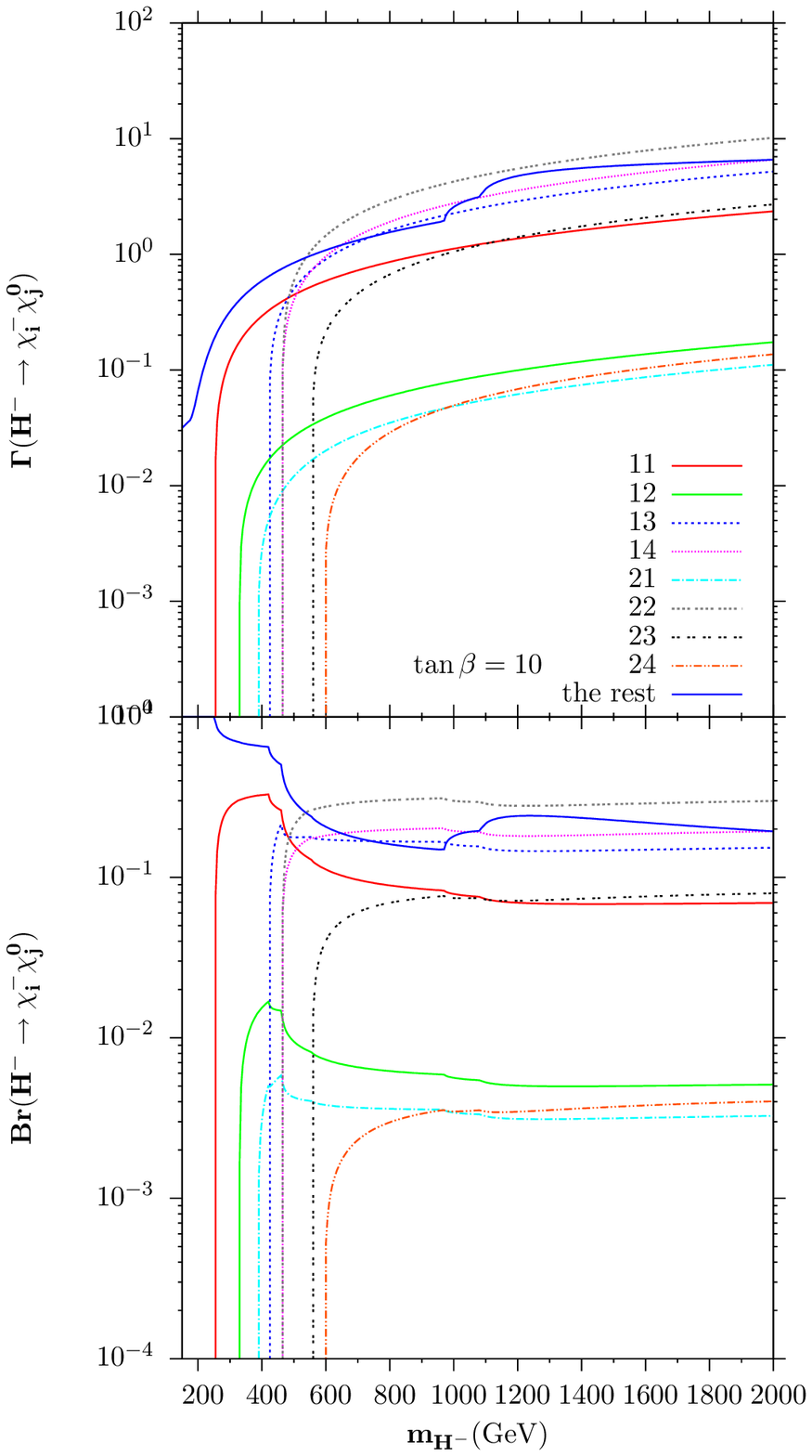} &
	\includegraphics[width=3.75in]{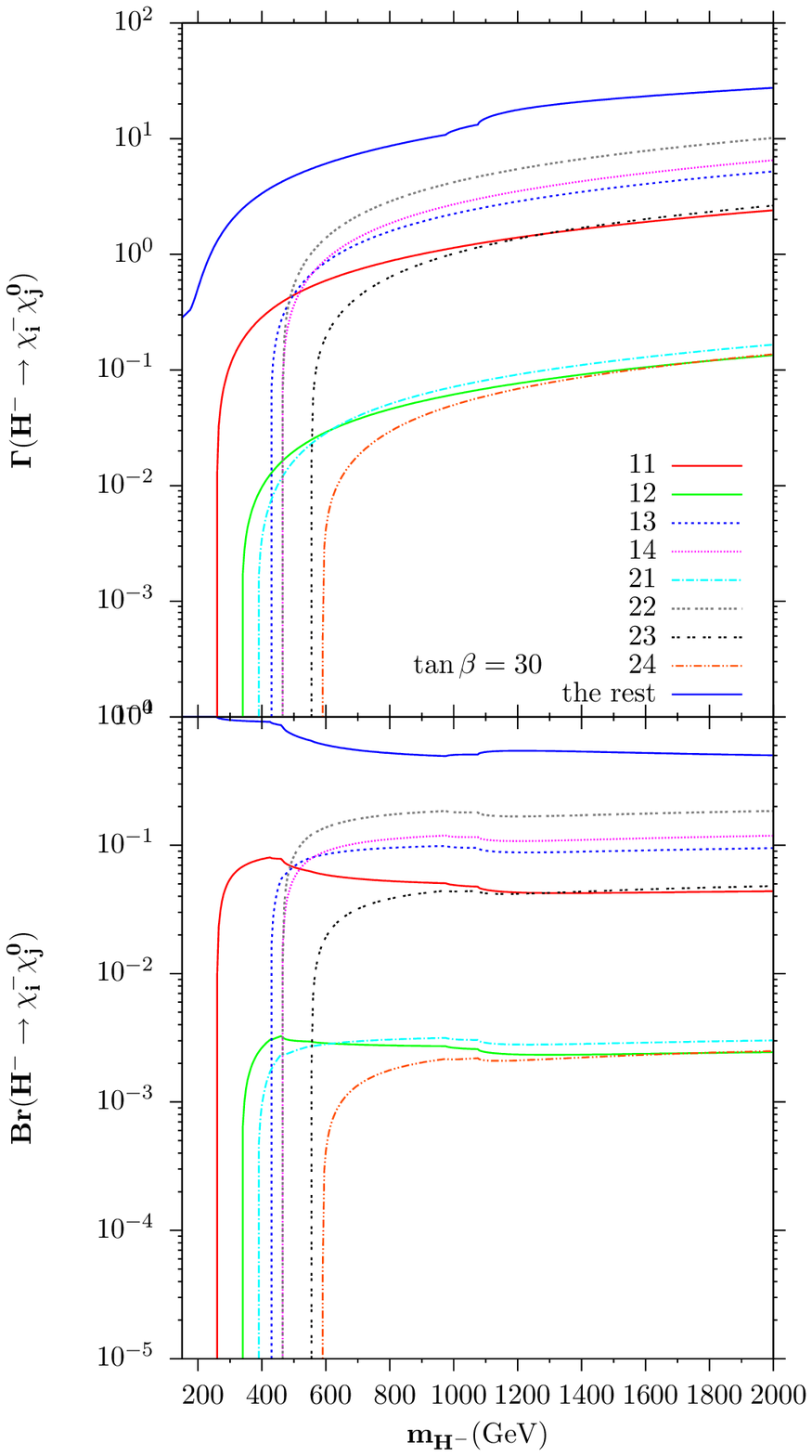}
	\end{array}$
\end{center}
\vskip -0.2in
      \caption{\underline{On the left:} $M_{SUSY}=500$ GeV, $\mu = 250$ GeV,  $\tan\beta = 10$, $M_2 = 200$ GeV, and $|A_{t,b,\tau}|=400$ GeV. \underline{On the right:} Same as on the left but for $\tan\beta = 30$.}
\label{fig:Br}
\end{figure}

As we approximate the partial decay widths of the channels by their tree level contribution in the denominator of Eq.~(\ref{ACP}), the denominator becomes $2\,\Gamma^0(\gen{-}{i}{j})$ where $\Gamma^0$ represents the tree level partial width. The partial decay widths of the decay into all possible eight channels of chargino-neutralino pairs with tree level approximation are shown in Fig.~\ref{fig:Br} as a function of the Higgs mass $m_{H^-}$, for some values of the free parameters of MSSM (the common scalar mass scale $M_{\rm SUSY}$, the gaugino mass parameter $M_2$, the trilinear coupling $A_f$, $\mu$, and $\tan\beta$). The decay widths are plotted on the left for intermediate $\tan\beta$ value, while the ones on the right are for large $\tan\beta$. For the calculation of {\it the rest} of charged Higgs channels, the {\tt FeynHiggs} program \cite{FeynHiggs} is used. Among these, there are fermion channels; $H^-\to f_i\nu_f$ among which $\tau\nu_\tau$ is the main decay mode for low Higgs mass region ($m_{H^-}\le 140$ GeV) and $\bar{t}b$ is for intermediate Higgs mass values ($m_{H^-}\ge m_t$), the Higgs-vector boson channels $H^\pm\to h^0W,\,H^0W,\,A^0W$, and the sfermion channels $H^\pm\to \tilde{f}_i\tilde{f}_j\;i,j=1...3$ where they are usually significant for heavy Higgs mass region. 

As seen from Fig.~\ref{fig:Br}, at intermediate $\tan\beta$, some of chargino-neutralino channels start being competitive and dominant with respect to the sum of the other open channels for a Higgs mass around $m_{H^-}\geqslant 600$ GeV. This is not the case at large $\tan\beta$ (the case for low $\tan \beta$ is similar to the case for high $\tan \beta$ and we do not graph it separately).  In these cases, in the heavy Higgs mass region, the eight chargino-neutralino channels split into two groups; \gen{-}{1}{2}, \gen{-}{2}{1}, and \gen{-}{2}{4} with the branching ratios less than 0.006 and the other five channels with the $\rm Br's$ around 0.1-0.3. Since usually a smaller branching ratio is an indication of larger observable CP asymmetries \cite{Atwood:2000tu}, we could expect that \gen{-}{1}{2}, \gen{-}{2}{1}, and \gen{-}{2}{4} may yield observable CP asymmetries. Of course an extensive numerical analysis is required and in the rest of the section  we investigate this by scanning the parameter space. 
\begin{figure}[h]
\begin{center}$
\hspace*{-1.2cm}
	\begin{array}{c@{\hspace{-0.5cm}}c}
	\includegraphics[width=3.5in]{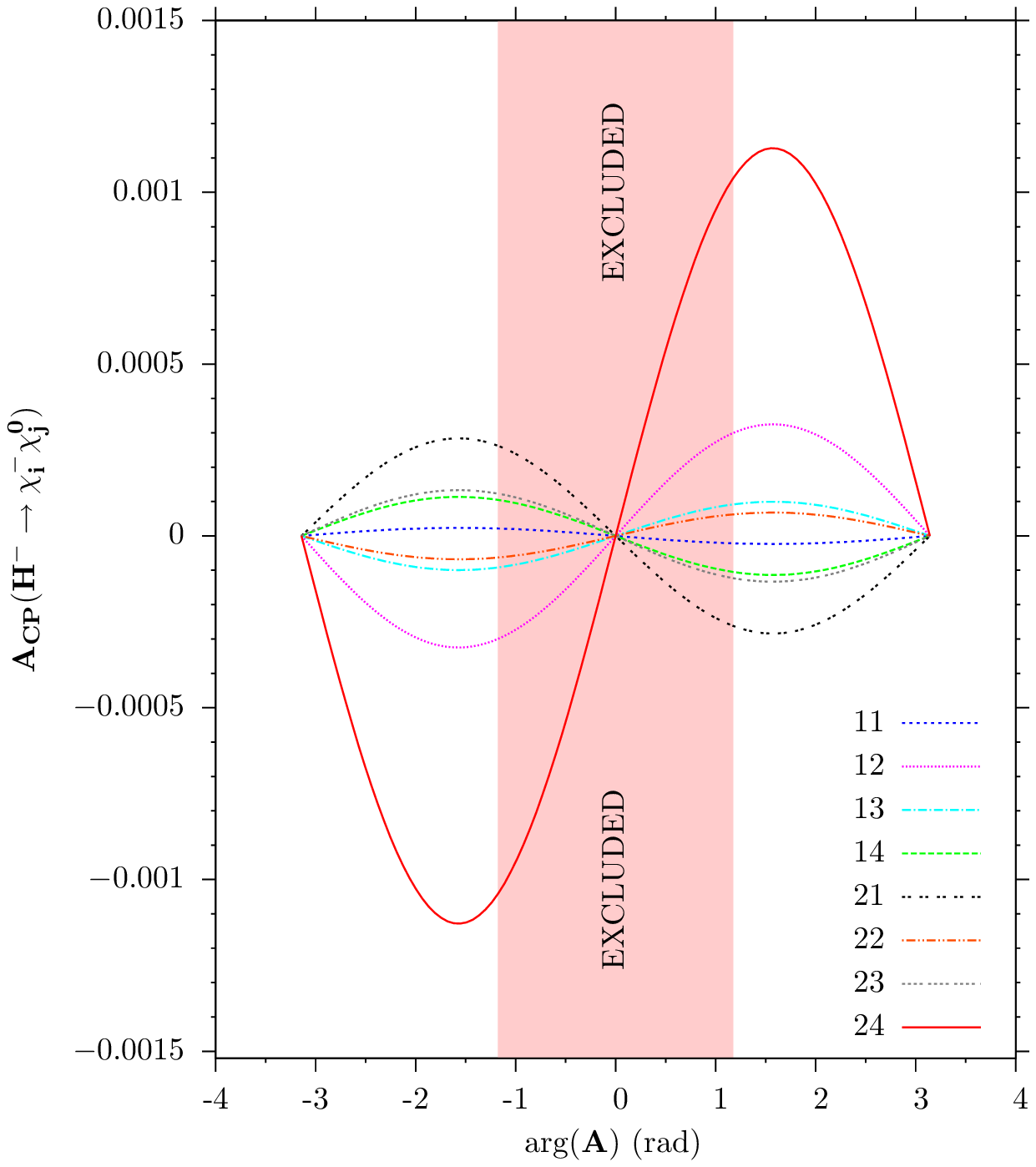} &
	\includegraphics[width=3.55in]{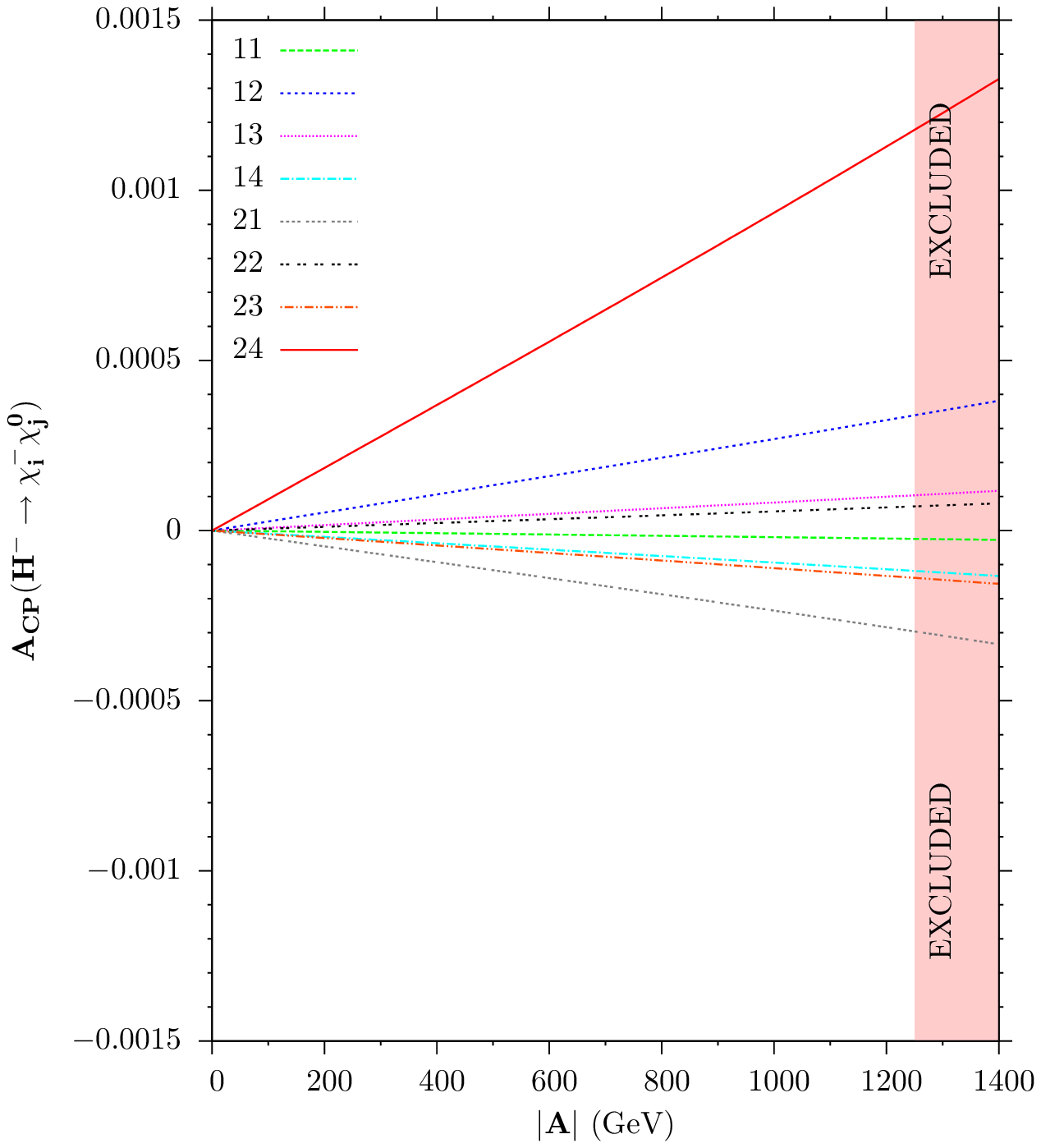} 
	\end{array}$
\end{center}
\vskip -0.2in
      \caption{\underline{On the left:} $M_{SUSY}=1$ TeV, $\mu = 250$ GeV,  $\tan\beta = 10$, $M_2 = 200$ GeV, $|A_{t,b,\tau}|=1.2$ TeV, $\arg(A_b)=\arg(A_\tau)=\arg(A_t)=\arg(A)$ and $m_{H^-} = 1$ TeV. \underline{On the right:} Same as on the left but $\arg(A)=\pi/2$. Excluded regions are disfavored by all experimental bounds considered here.}
\label{fig:phiA-A}
\end{figure}

We first check the sensitivity of the CP asymmetry to the free parameters and phases and then scan the sensitive parameters for the maximal asymmetry on a multidimensional parameter space. At each stage we show both the regions allowed by imposing only supersymmetric mass limits, and then show the parameter regions excluded by imposing the experimental constraints coming from $b \to s \gamma$, $(g-2)_{\mu}$, $\Delta\rho$ and EDMs. The current experimental mass lower bounds  considered are $m_{\tilde{u},\tilde{\tau}}> 96  {\rm ~GeV},\; m_{\tilde{d}} > 89  {\rm ~GeV},\; m_{\chi^0} > 50  {\rm ~GeV}$, and  $m_{\chi^+} > 94  {\rm ~GeV}$ \cite{Yao:2006px} .

The constraints we included to restrict the CP asymmetry are: $(g-2)_{\mu} \le 40\times10^{-10}$ \cite{Bennett:2004pv}
 and  $|\Delta\rho|\le 10^{-3}$ \cite{Yao:2006px}; as well as electric dipole moment constraints: $|d_{\rm n}| < 6\times10^{-26}\;\,e.cm$ for neutron \cite{Yao:2006px}, $|d_{\rm Th}| < 1.6\times10^{-27}\;\,e.cm$ for thallium\footnote{The upper bound is indeed the translated one on the electron EDM $d_e$.}  \cite{Regan:2002ta}, and $|d_{\rm Hg}| < 2\times10^{-28}\;\,e.cm$ for mercury \cite{Romalis:2000mg}. For  $b \to s \gamma$ we use at $3\sigma$ \cite{bsgamma}
\begin{eqnarray}
2.53 \times 10^{-4} < Br ( b \to s \gamma ) < 4.34 \times 10^{-4}.
\label{boundbsg}
\end{eqnarray}
The calculation and inclusion of these bounds have been done with the use of {\tt FeynHiggs} \cite{FeynHiggs}. We don't impose SUSY GUT relations for gaugino masses. Since the $b\to s \gamma$  decay can be accommodated by a rather heavy gluino mass, and this does not influence our results for the CP asymmetry, this assumption allows us   to keep $M_2$ rather light. So, we assume a gluino mass of $m_{\tilde g}= 5$ TeV but express $M_1$ in terms of $M_2$  using GUT relations. We find that inclusion of the experimental constraints not only restricts the parameter space but also has a significant effect on the CP asymmetry overall.

\begin{figure}[htb]
\begin{center}$
\hspace*{-1.2cm}
	\begin{array}{c@{\hspace{-0.5cm}}c}
	\includegraphics[width=3.61in]{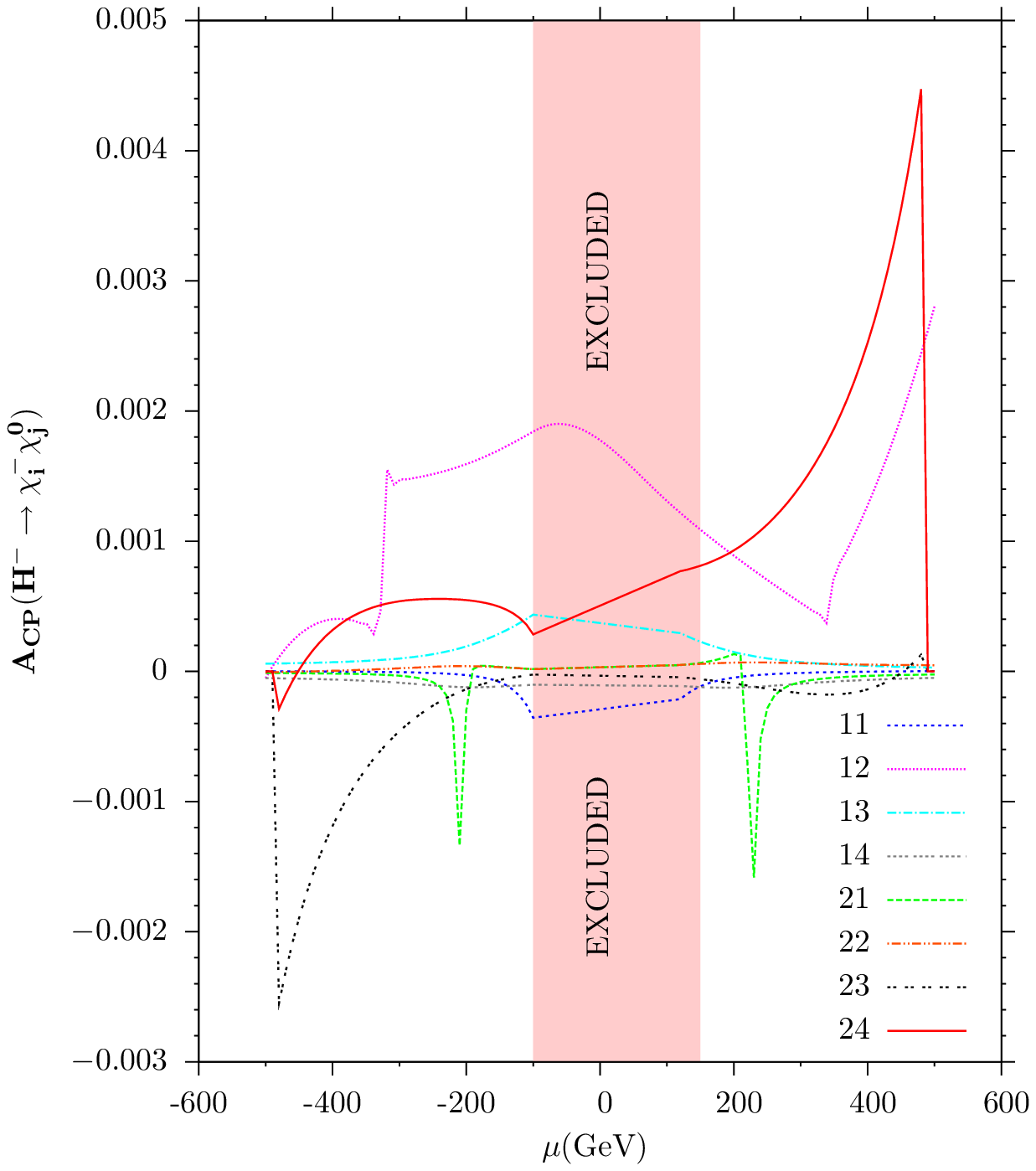} &
	\includegraphics[width=3.5in]{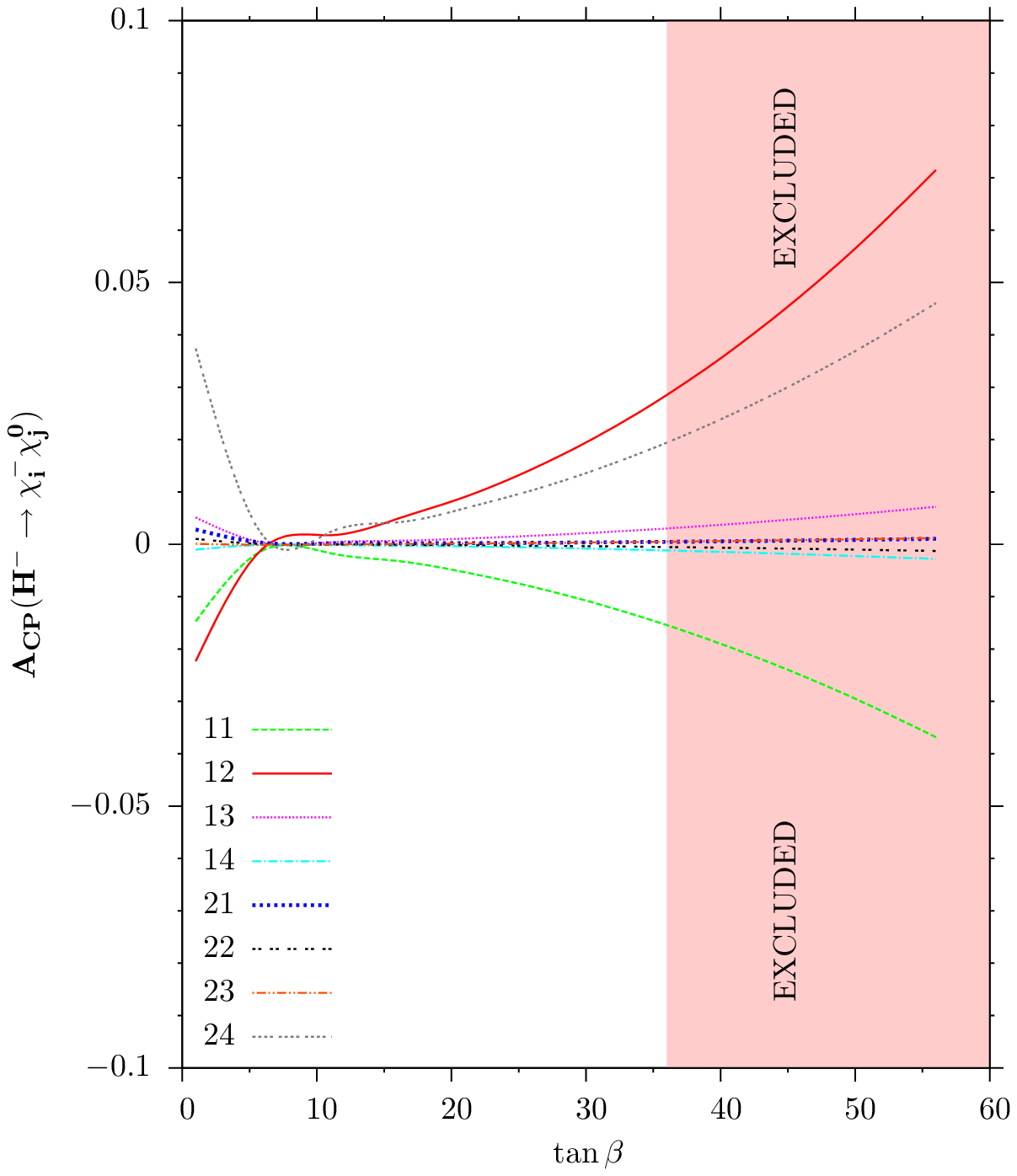} 
	\end{array}$
\end{center}
\vskip -0.2in
      \caption{\underline{On the left:} $M_{SUSY}=1$ TeV, $\tan\beta = 10$, $M_2 = 200$ GeV, $|A_{t,b,\tau}|=1.2$ TeV, $\arg(A_b)=\arg(A_\tau)=\arg(A_t)=\pi/2$ and $m_{H^-} = 1$ TeV. Excluded region is disfavored by mass lower bounds. \underline{On the right:} Same as on the left but $\mu = 200$ GeV. Excluded region is disfavored by all experimental bounds considered here.}
\label{fig:MUE-TB}
\end{figure} 
In Fig.~\ref{fig:phiA-A}, the CP asymmetry is depicted as a function of a common trilinear coupling $A$ for the third generation lepton and quarks, assuming they share a common phase, $\arg(A)$. The Higgs mass is assumed rather heavy, 1 TeV. The maximal CP asymmetry is obtained at $\arg(A) =\pi/2$ which is still allowed after excluding the experimentally disfavored region.  The largest asymmetry is obtained for the channels \gen{-}{2}{4}, \gen{-}{1}{2}, and \gen{-}{2}{1}. As expected the larger $|A|$ is, the bigger the asymmetry becomes. In these figures, almost all the contribution to the asymmetry comes from the interference of the vertex diagram $\#\, 7$ in Fig.~\ref{fig:relevant} with tree level. The asymmetry at maximum is around 0.1\% for \gen{-}{2}{4} and much smaller for the rest. Although we did not include it in a separate figure, we also analyzed the CP asymmetry as a function of the phase of $A_b$ by assuming the other phases zero, and separately for the phase of $A_\tau$ and found no significant effect. Assuming no phases from $A_b$ and $A_\tau$ leads to no contributions to $A_{CP}$ from the diagrams $\#\, 3, 6, 8, 9$, and $10$. This has been verified numerically.
\begin{figure}[htb]
\begin{center}$
\hspace*{-1.2cm}
	\begin{array}{c@{\hspace{-0.5cm}}c}
	\includegraphics[width=3.5in]{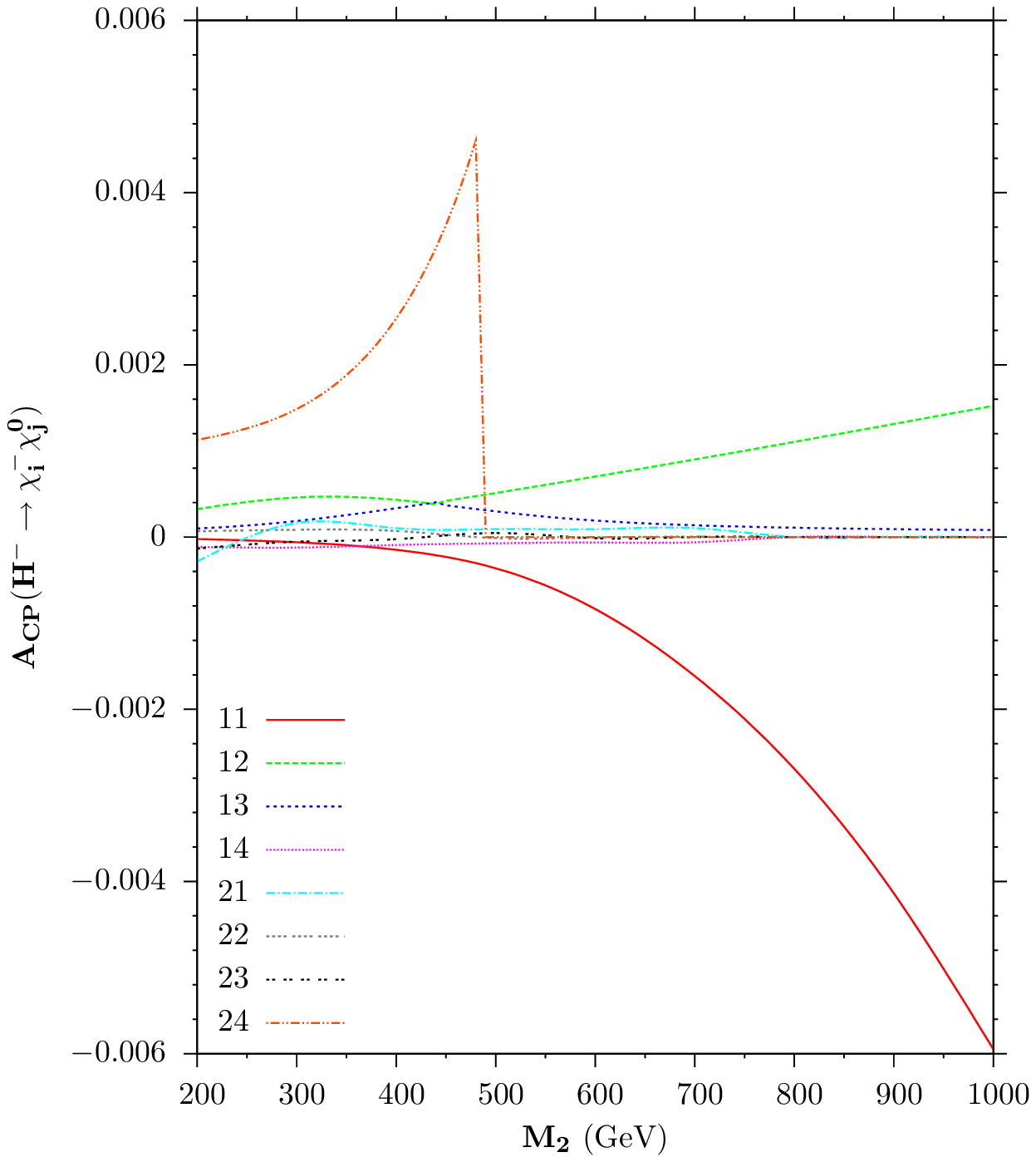} &
	\includegraphics[width=3.5in]{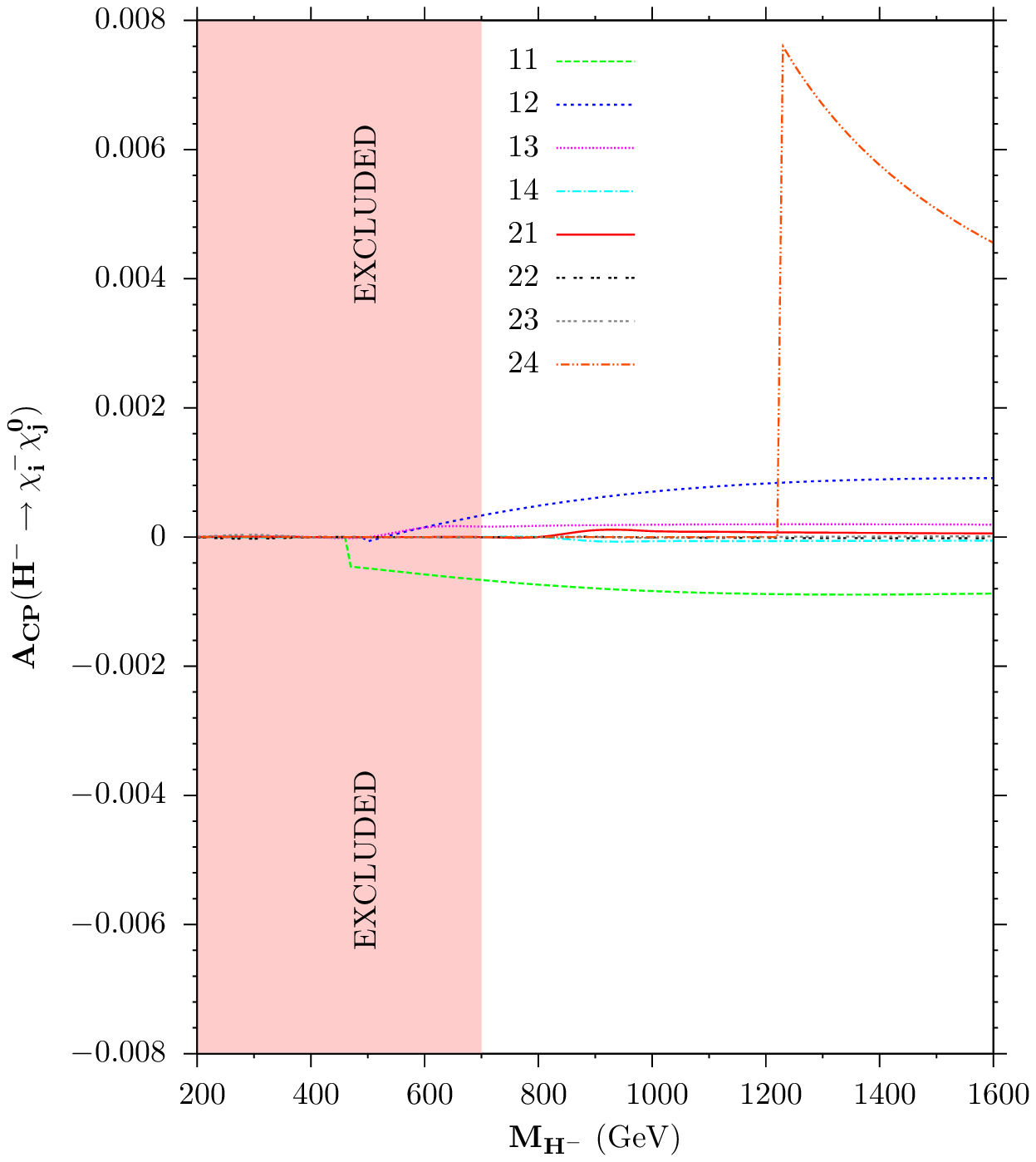} \\
	\includegraphics[width=3.5in]{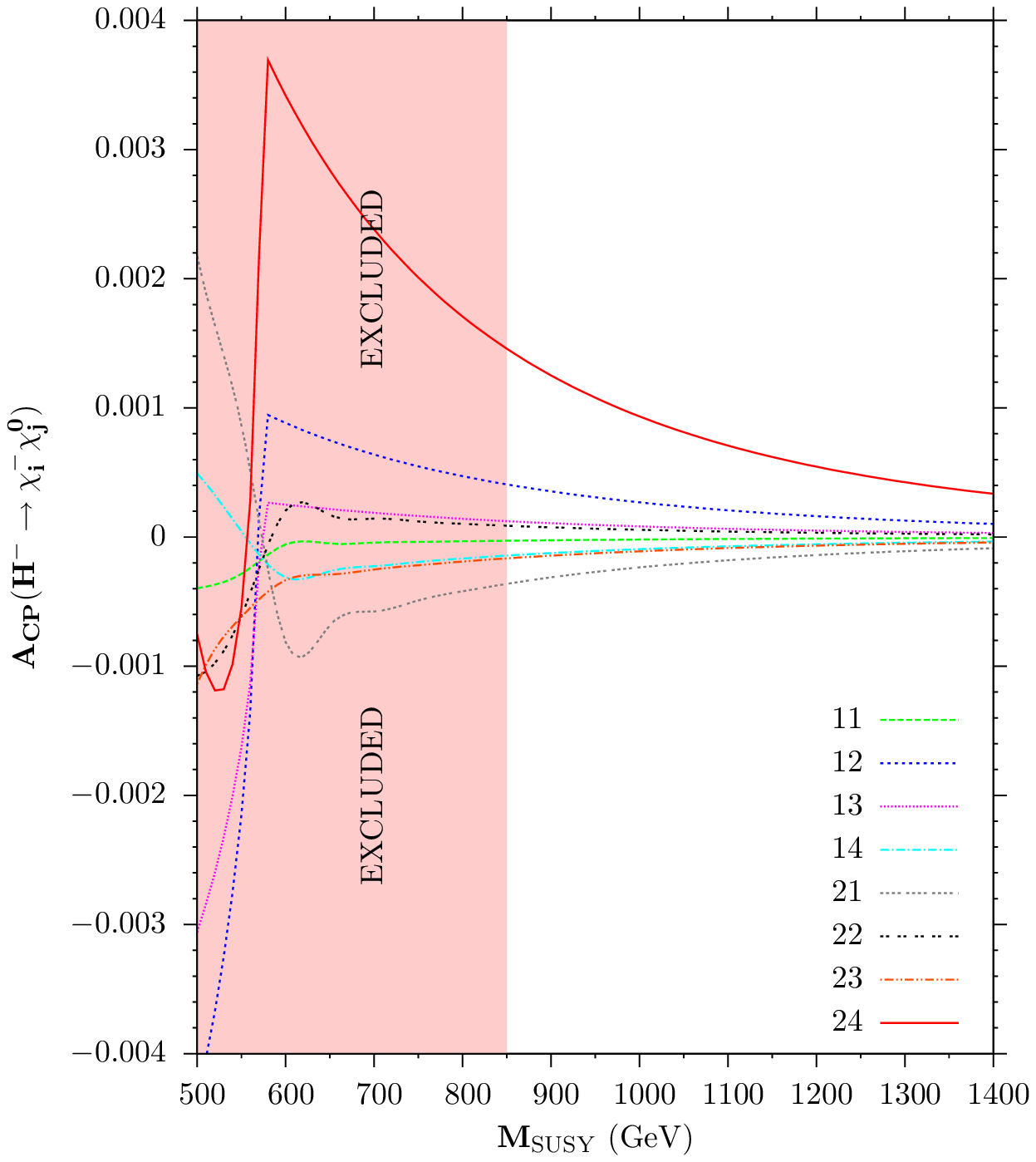}
	\end{array}$
\end{center}
\vskip -0.2in
      \caption{\underline{For the first row:} $M_{SUSY}=1$ TeV, $\tan\beta = 10$, $|A_{t,b,\tau}|=1.2$ TeV, $\arg(A_b)=\arg(A_\tau)=\arg(A_t)=\pi/2$. \underline{\it Left:} $m_{H^-} = 1$ TeV. \underline{\it Right:} $M_2 = 200$ GeV. \underline{For the second row:} Same as above but $|A_{t,b,\tau}|=1$ TeV. Excluded regions are disfavored by all experimental bounds considered here.}
\label{fig:M2MHpMSusy}
\end{figure}  

In the next figure, Fig.~\ref{fig:MUE-TB}, as we vary $\mu$ in the $(-500, 500)$ GeV range,  the magnitude of the CP asymmetry becomes as large as $(0.3-0.5)\%$ for \gen{-}{1}{2}, \gen{-}{2}{4}, and \gen{-}{2}{3}. This time the region excluded violates the experimental lower bounds on the lightest chargino and neutralino mentioned above\footnote{The boundary points of the excluded region vary with respect to each decay mode. In Fig.~\ref{fig:MUE-TB}, we consider some average values for all of them.}. The asymmetry is sensitive with respect to variations in  $\tan\beta$ and reaches $(3-4)\%$ in the allowed region for (again) \gen{-}{1}{2}. 

The next figure,  Fig.~\ref{fig:M2MHpMSusy}, presents the asymmetry as a function of $M_2, m_{H^-}$, and $M_{\rm SUSY}$. The asymmetry vanishes for all channels except \gen{-}{1}{1} and \gen{-}{1}{2} for larger $M_2$ values for kinematical reasons (the heavier channels cannot be produced in the decay).  As a function of the Higgs mass, the CP asymmetries for \gen{-}{1}{2} and \gen{-}{2}{4} are the most sensitive. The EDM constraints especially allow neither $M_{\rm SUSY}$ nor $m_{H^-}$ to be light. Choosing a smaller $A$ value with a smaller phase would relax this restriction significantly, but would also decrease the CP asymmetry. 

\begin{figure}[h]
\begin{center}$
\hspace*{-1.2cm}
	\begin{array}{c@{\hspace{-0.5cm}}c}
	\includegraphics[width=3.5in]{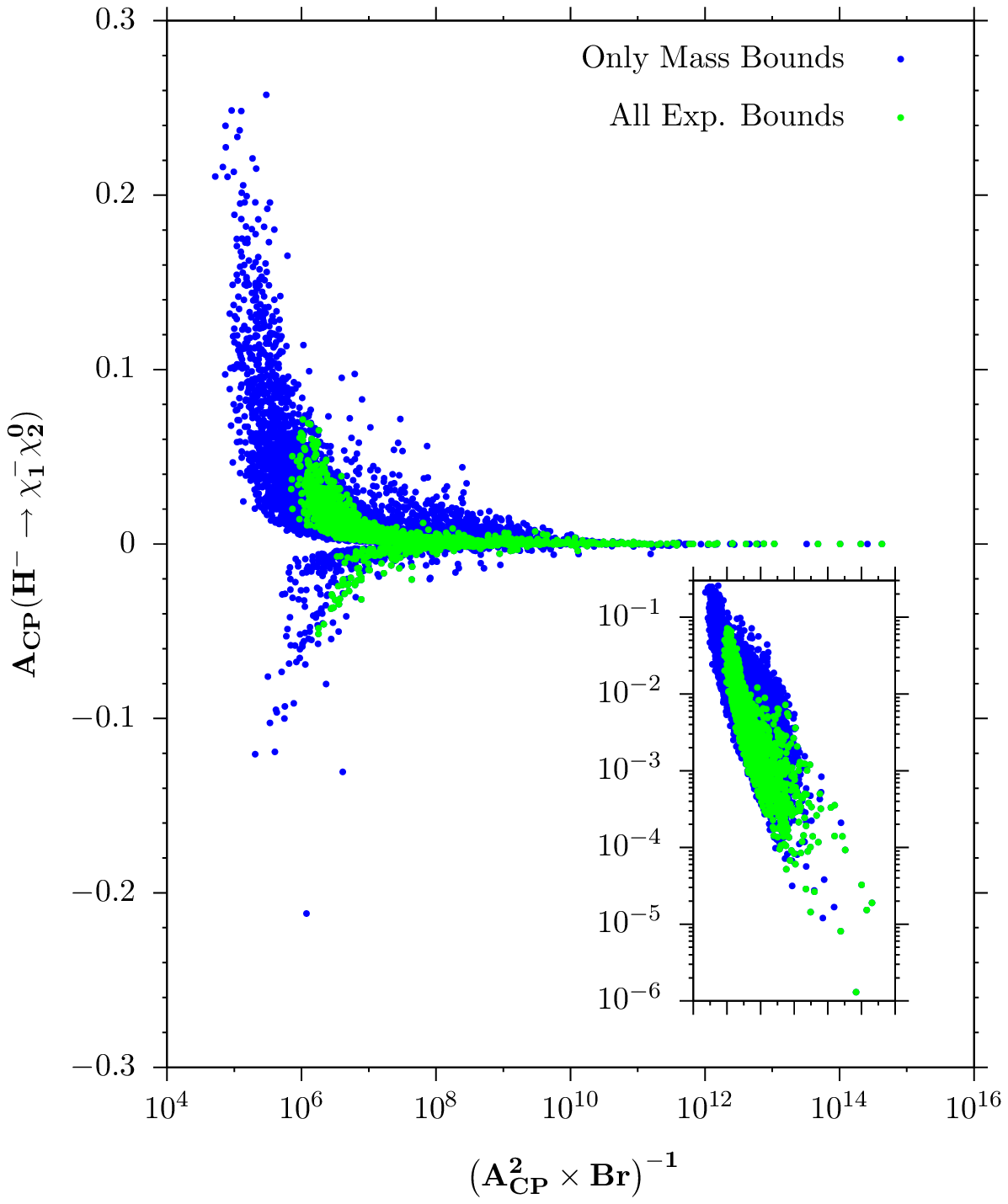} &
	\includegraphics[width=3.6in]{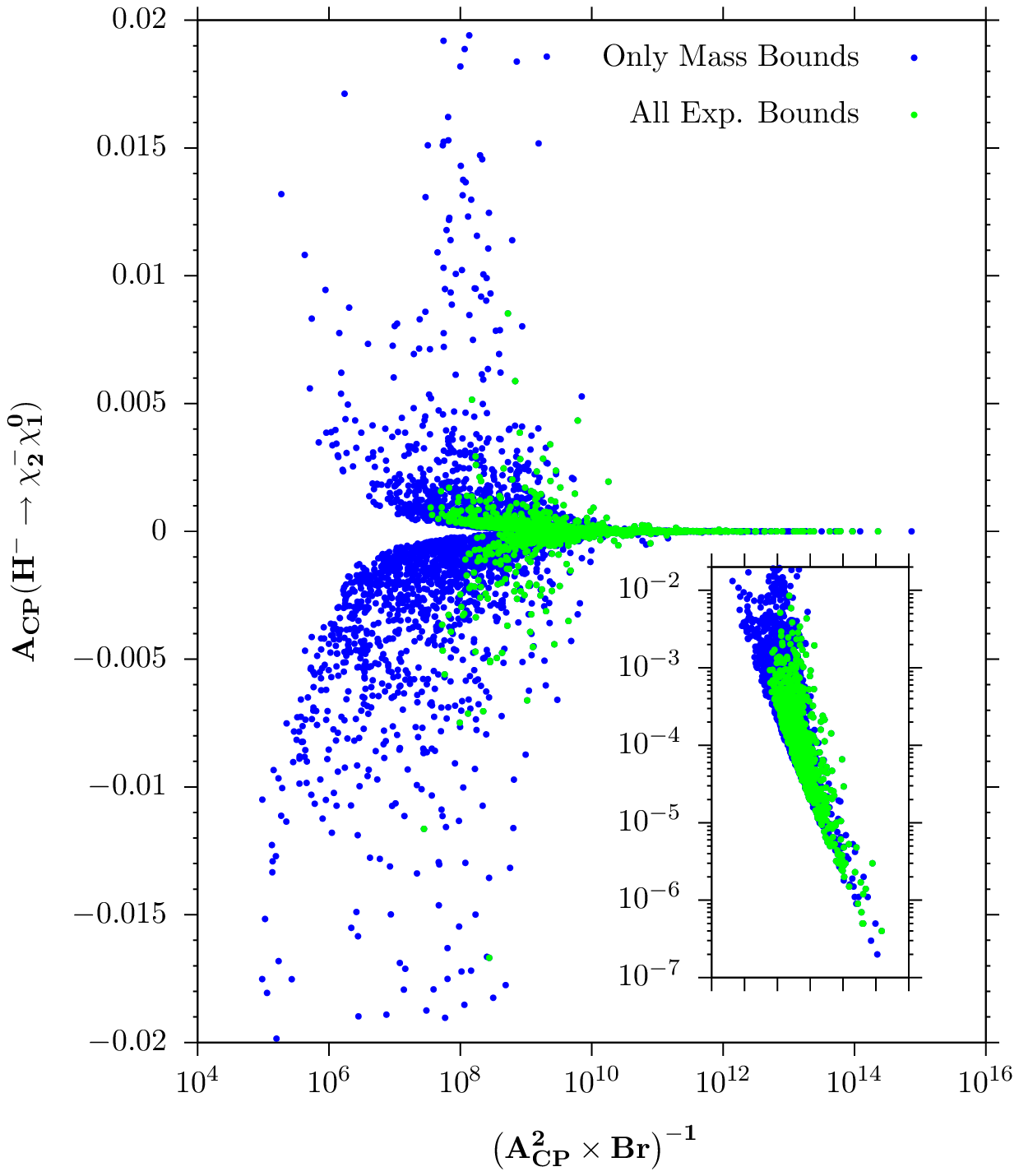} 
	\end{array}$
\end{center}
\vskip -0.2in
      \caption{The scatter plot for the charged Higgs decays $H^-\to \chi^-_{1(2)} \chi^0_{2(1)}$ in the $(A_{CP}^2\times Br)^{-1}-A_{CP}$ plane. These are obtained by scanning the sensitive parameters $A\in(0,1400)\,{\rm GeV},\, m_{H^-}\in(600,1500)\,{\rm GeV},\,\tan\beta\in(1,50),\, M_2\in(200,1000)\,{\rm GeV},\,\mu\in(-500,500)\,{\rm GeV}$, and $M_{\rm SUSY}\in(600,1200)\,{\rm GeV}$.  The phase is fixed, $\arg(A_t)=\pi/2$. The blue points satisfy only the mass bounds and the green ones satisfy all constraints. The x-axes are in the logarithmic scale. The small graphs inside each graph represent the positive asymmetry $A_{CP}$ in the logarithmic scale.}
\label{fig:scatter1221}
\end{figure}

\begin{figure}[htb]
\begin{center}$
\hspace*{-1.2cm}
	\begin{array}{c@{\hspace{-0.5cm}}c}
	\includegraphics[width=3.6in]{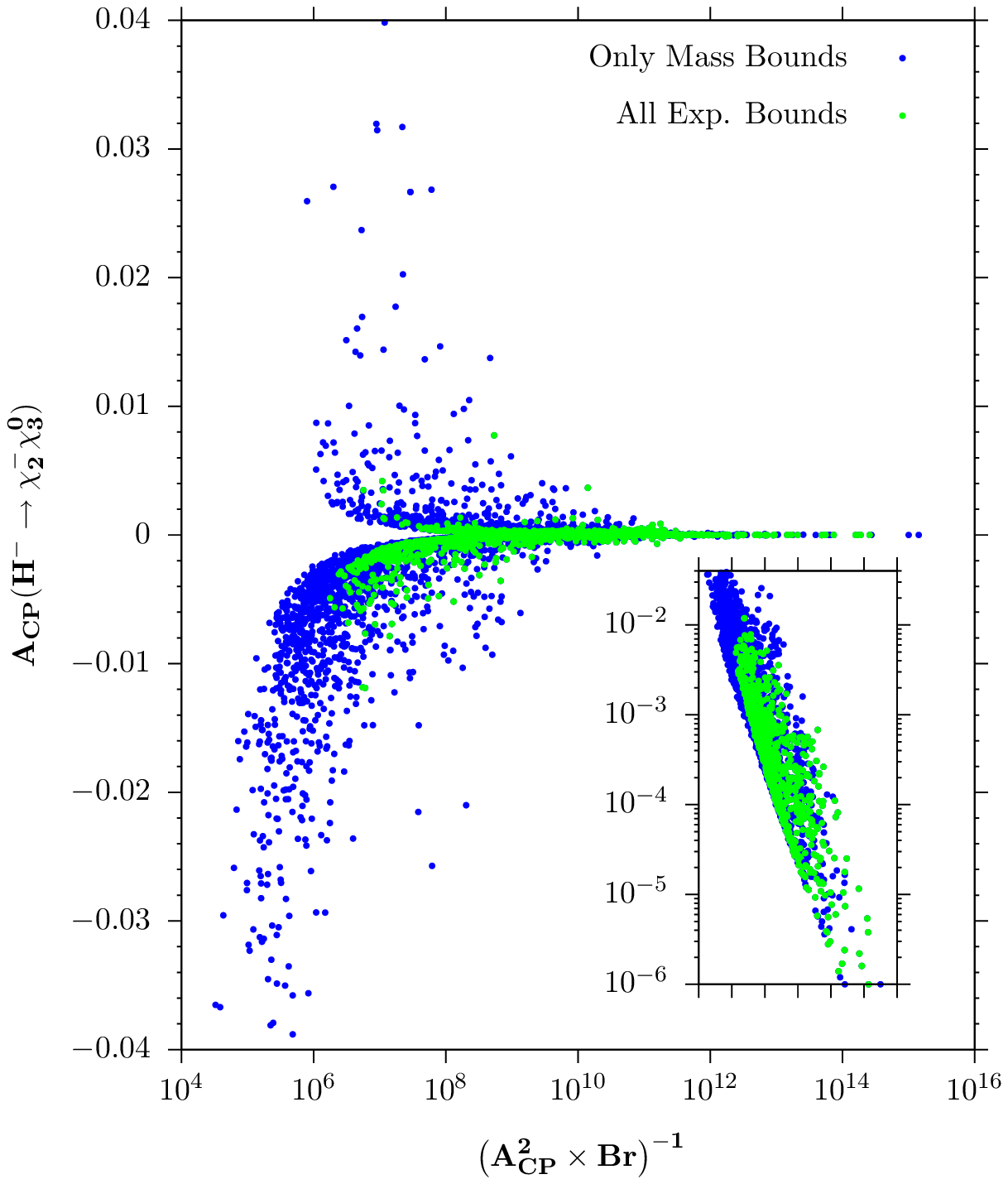} &
	\includegraphics[width=3.5in]{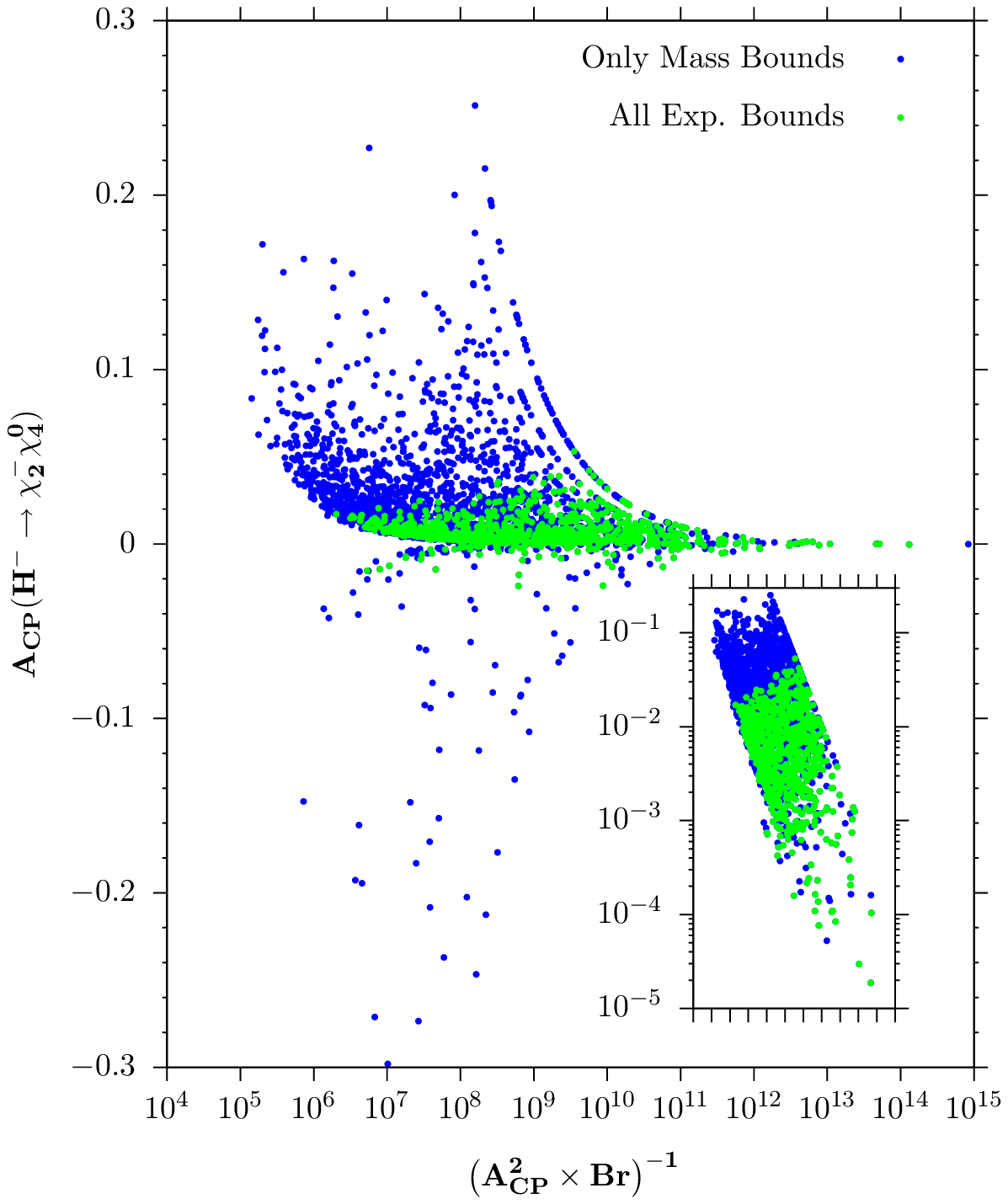}	
	\end{array}$
\end{center}
\vskip -0.2in
      \caption{The scatter plot for the charged Higgs decays $H^-\to \chi^-_{2} \chi^0_{3(4)}$ in the $(A_{CP}^2\times Br)^{-1}-A_{CP}$ plane. These are obtained by scanning the sensitive parameters $A\in(0,1400)\,{\rm GeV},\, m_{H^-}\in(600,1500)\,{\rm GeV},\,\tan\beta\in(1,50),\, M_2\in(200,1000)\,{\rm GeV},\,\mu\in(-500,500)\,{\rm GeV}$, and $M_{\rm SUSY}\in(600,1200)\,{\rm GeV}$.  The phase is fixed, $\arg(A_t)=\pi/2$. The blue points satisfy only the mass bounds and the green ones satisfy all constraints. The x-axes are in the logarithmic scale. The small graphs inside each graph represent the positive asymmetry $A_{CP}$ in the logarithmic scale. Thus, in the small panel for \gen{-}{2}{3}, $A_{CP}$ is multiplied by -1.}
\label{fig:scatter2324}
\end{figure}

The final part of the numerical analysis is devoted to the scan for maximal CP asymmetry among the four of eight channels (two light from Fig. \ref{fig:scatter1221} and two heavy ones from Fig. \ref{fig:scatter2324}) over sensitive variables of the parameter space ($M_{\rm SUSY}, m_{H^-}, M_2, \mu, \tan\beta$). We choose the following four channels; \gen{-}{1}{2}, \gen{-}{2}{1}, \gen{-}{2}{3}, and \gen{-}{2}{4}, which seem most promising to yield larger asymmetries. For the sake of efficiency, we included some of the results of the previous parameter dependence.  Since we verified that the effect of the phases $\arg(A_b)$ and  $\arg(A_\tau)$ are negligible, we set them to zero, while keeping the absolute value of $A_b$, since there are diagrams with scalar-up/scalar-down quark loops which are not negligible. We also fix the phase of $A_t$ as $\pi/2$, as we know that it maximizes the CP asymmetry.  The absolute values of $A_t$  and $A_b$ are chosen as $1.2$ TeV, which appear to be values in the region which maximizes the asymmetry. Then we vary the rest of the parameters in the following ranges; $M_{\rm SUSY} \in(600,1200)\,$ GeV, $m_{H^-}\in(600,1500)\,$GeV, $M_2\in(200,1000)\,$ GeV, $\mu\in(-500,500)\,$ GeV and $\tan\beta \in(2,50)$.  

In Figs.~\ref{fig:scatter1221} and \ref{fig:scatter2324}, we present the scatter plots of the modes  \gen{-}{1}{2}, \gen{-}{2}{1}, \gen{-}{2}{3}, and \gen{-}{2}{4} in the $(A_{CP}^2\times Br)^{-1}-A_{CP}$ plane. As before \cite{Frank:2007ca}, we introduce here the quantity $(A_{CP}^2\times Br)^{-1}$ as a measure of the number of required charged Higgs bosons for the observibility of the asymmetry. This is an order of magnitude estimation and the exact value can be obtained by multiplying it with the factor $s^2/\epsilon$ where $s$ is the standard deviation and $\epsilon$ is the detection efficiency. There are two types of points on the scan, color-coded as blue and green. The blue dots only obey the basic mass lower bounds. As one can see, under these requirements only, each decay mode can generate a CP asymmetry as large as $30\%$\footnote{The mode \gen{-}{2}{1} can also have blue points as large as $30\%$ but we restricted the extent of the vertical axis for better resolution.}.  Of course the input set for the parameters is different for each channel.  These are listed for the maximal asymmetry points in Table~\ref{Table}. The number of charged Higgs bosons needed to observe the predicted CP asymmetries is in the range of $(10^5-10^9)\times s^2/\epsilon$. In the small panels in Figs.~\ref{fig:scatter1221} and \ref{fig:scatter2324} we also showed the distribution of points in logarithmic $A_{CP}$ scale. 

However, if in addition to the mass bounds, one adds the experimental constraints on $b \to s \gamma$, $(g-2)_{\mu}$, $\Delta\rho$ and EDMs, only the green points on the graphs survive and the maximal asymmetries for each channel go down to few percent level, with the requirement of a larger number of Higgs to be produced for better statistics. Again, the maximal points and the corresponding input set and the required Higgs bosons are given in Table~\ref{Table} as separate columns for comparison. As seen from the Table, for maximal asymmetry all four channels favor a heavy charged Higgs mass greater than 1 TeV and also either large or small $\tan\beta$. In the case where we include only the mass bounds,  the low $M_{\rm SUSY}$ values in the chosen range $(600, 1200)$ GeV are favored since in such cases $m_{H^-}>m_{\tilde{u}_i}+m_{\tilde{d}_j}$ is satisfied, the squarks can go on-shell in the loop and contribute to $A_{CP}$ significantly\footnote{Lighter charged Higgs contribute dominantly to $b\to s\gamma$ and to satisfy that the constraint within the 3 $\sigma$ level one has to consider the non-minimal flavor violating MSSM scenarios. See Ref.~\cite{Cao:2007dk} for the details.}. However, the experimental constraints, especially EDMs, require a heavier SUSY scale,  around 1 TeV. Note that we set the CP phase $\arg(A_t)=\pi/2$  for $A_t$ of 1.2 TeV magnitude. For these values, mass parameters in the top scalar quark system affect the light quark EDMs (through chargino loops) and lighter SUSY scales are excluded to satisfy the EDM bounds.
\begin{table}[t]
	\caption{The maximal CP asymmetries and the number of required charged Higgs bosons for the four promising channels, \gen{-}{1}{2}, \gen{-}{2}{1}, \gen{-}{2}{3}, and \gen{-}{2}{4}. We also include the input parameter set for each channel. We show the results by including all experimental constraints and compare with the results where only mass bounds are consired.} \label{Table}     
\vskip 0.08in
\begin{minipage}{1.01\textwidth}
\begin{tabular}{|c||c|c||c|c|c|c|c||c|} 
\hline 
Chan.&
$A_{CP}$(\%)&
$(A_{CP}^{2}\times Br)^{-1}(\frac{s^{2}}{\epsilon})$&
$m_{H^{-}}$(GeV)&
$M_{{\rm SUSY}}$(GeV)&
$M_{2}$(GeV)&
$\mu$(GeV)&
$\tan\beta$&
Consts.\\
\hline
\hline 
12&
-36.7&
$3\times10^{5}$&
1040&
-600&
680&
-500&
41&
Mass Only\\
\hline 
12&
7.1&
$1\times10^{6}$&
1320&
1040&
840&
-100&
44&
All\\
\hline
\hline 
21&
-24.5&
$1\times10^{6}$&
1280&
640&
560&
460&
29&
Mass Only\\
\hline 
21&
-2.3&
$1\times10^{8}$&
1440&
1000&
200&
220&
38&
All\\
\hline
\hline 
23&
-17.5&
$3\times10^{4}$&
1120&
600&
240&
-460&
41&
Mass Only\\
\hline 
23&
-1.2&
$6\times10^{6}$&
1160&
960&
200&
-500&
2&
All\\
\hline
\hline 
24&
-57.6&
$2\times10^{9}$&
1480&
760&
720&
-180&
14&
Mass Only\\
\hline
24&
8.8&
$5\times10^{10}$&
1500&
960&
680&
-220&
5&
All\\
\hline
\end{tabular}
	\end{minipage}
\end{table}

The most promising channel for observing a large asymmetry is \gen{-}{1}{2}, which could show a maximal CP asymmetry around $7\%$ with approximately $10^6\times s^2/\epsilon$ required charged Higgs produced at colliders.  Even though  \gen{-}{2}{4} can produce a slightly larger asymmetry, it is not a better channel since it not only requires more Higgs particles, but also it is a decay channel into much heavier states. The maximal asymmetry for \gen{-}{2}{4} requires $10^{11}\times s^2/\epsilon$ events, much more than  for the other three modes. Moreover, its branching ratio is of the order of $10^{-9}$. Thus  the CP asymmetry in \gen{-}{1}{2} has a better chance to be observed than the other three. Its branching ratio is around $2\times 10^{-4}$ which is small, but it is five orders of magnitude bigger than that of \gen{-}{2}{4}.  Note that the distribution of the points with respect to zero axis is not symmetric and the CP asymmetry is mostly positive especially for the green points which satisfy all constraints (except \gen{-}{2}{3}). Though the relative sign is a prediction of MSSM, the absolute sign is a result of choosing the phase of $A_t$ to be positive. 

\section{Summary and Conclusion}
\label{conc}
We studied the CP asymmetry of the charged Higgs decays into pairs of charginos and neutralinos in the framework of the MSSM. We neglected  the possibility of flavor violation in the squark sector.   We first analyzed the partial decay widths and the corresponding branching ratios for these channels. For a region of the parameter space with intermediate values of $\tan\beta$ and for $m_{H^-}\geqslant 600$ GeV some chargino-neutralino channels are competitive with, and sometimes dominant over, other open channels.  We then investigated the size of the asymmetry for various final decay states and the likelihood of observing it at colliders.  A non-zero CP asymmetry requires an absorptive phase, for which we considered possible interference terms between tree-level and one-loop diagrams. We calculated the imaginary parts and presented both analytical expressions for the discontinuity in the diagrams with the help of Cutkosky rules, as well as numerical results. We have shown that the dominant phase is the one describing the left-right mixing the top scalar quark  mass matrix and the asymmetry is maximum for $\arg(A_t)=\pi/2$, while $\arg(A_b)$ and $\arg(A_\tau)$ give negligible contributions. We analyzed the dependence of the asymmetry on the parameters of MSSM: $M_{SUSY},\,\mu,\,\tan\beta,\,M_2 ,\,|A_{t,b,\tau}|,\, \arg(A_{t,b,\tau})$ and $m_{H^-}$. In each numerical and graphical investigation we include bounds from  $b \to s \gamma$, $(g-2)_{\mu}$, $\Delta\rho$ and the EDMs.

We then performed a scan of the significant parameter space for maximizing the asymmetry, and correlate it  to the number of charged Higgs needed to see the corresponding asymmetry shown to be proportional to $(A_{CP}^2\times Br)^{-1}$. In the scan we included the experimental bounds on the masses of supersymmetric particles only, as well as the previously mentioned low energy bounds. We first note that the inclusion of mass bounds restricts the asymmetry slightly, while the other bounds pose more severe restrictions on the CP asymmetries.  In particular, the EDMs, thought to restrict parameters in the up and down scalar quark only, impose severe restrictions even on the masses and mixing of the third family of scalar quarks, confirming the expectation that the scalar quark mass parameters must be in the TeV region. Heavy scalar quark masses and heavy gluino masses are also needed to satisfy the $b\to s \gamma$ constraint, while the bounds on $(g-2)_\mu$ and $\Delta\rho$ are not as stringent. Our analysis underscores the importance of including these constraints on the parameter space, like in the analysis of top quark  production and decays \cite{Cao:2007dk}.

For the allowed regions, the most promising channel appears to be \gen{\pm}{1}{2}, where both the asymmetry can reach 7\% and the number of charged Higgs required to observe them is yet the smallest, of order $10^6\times s^2/\epsilon$. This channel may be competitive with the quark channels \cite{Frank:2007ca}. Of course, this is encouraging, as light charginos and neutralinos are most likely to be produced in charged Higgs decays in the first place. By comparison, the other channels are less promising. For instance, for heavy chargino-neutralino channel, \gen{-}{2}{4}, almost $10^{11}\times s^2/\epsilon$ charged Higgs events (with a branching ratio of the order of $10^{-9}$) are required for detecting an asymmetry of (8 - 9)\%.

\section{Acknowledgment}
This work is supported in part by NSERC of Canada under the Grant No. SAP01105354.

\appendix*
\section{The Method - Calculation of the Imaginary Part}
We calculate the absorptive part of the loop diagrams by applying the Cutkosky rules. In general, in the loop integration, we end up with a numerator with scalar, vectorial, or tensorial structures (and their pseudo counterparts), depending on the types of particles running in the loop. For a more complete discussion of our method and the cut results for the two-point (self-energy) case and three-point (vertex-type) scalar case, please see \cite{Frank:2007ca}.  We give here only the final formulas for the three-point (vertex-type) vectorial and tensorial cases with the horizontal cut:
\begin{eqnarray}
\displaystyle
 \!\!\!\Delta C^\mu\!\! &=&\!\! \frac{2\pi i\, \Theta(p_3^2-(m_1+m_3)^2)}{\sqrt{\lambda(p_1^2,p_2^2,p_3^2)}}\,\left \{\log\left(\frac{\alpha_3+\beta_3}{\alpha_3-\beta_3}\right)p_1^\mu
 -\left[(t-\alpha_3 v)\,\log \left(\frac{\alpha_3+\beta_3}{\alpha_3-\beta_3}\right) \right. \right. \nonumber \\
 &+&\left. \left.\frac{2\beta_3}{\alpha_3-\beta_3}\left (t-\beta_3v+\frac{2p_3^2}{\lambda}(\alpha+4p_1^2t)
  \right )\right]\frac{p_2^\mu}{2\,p_3^2}\right \},\nonumber\\
\!\!\!\Delta C^{\mu\nu}\!\! &=&\!\! \frac{2\pi i\, \Theta(p_3^2-(m_1+m_3)^2)}{ \sqrt{\lambda(p_1^2,p_2^2,p_3^2)}}\,\left[ \frac{g^{\mu\nu}}{2p_3^2} \Delta C_{3g} + p_1^\mu\, p_1^\nu\,  \!\Delta C_{p_1} + \frac{(p_1^\mu\, p_3^\nu+p_1^\nu\, p_3^\mu)}{2p_3^2 }\,\Delta C_{p_1p_3} \right.\nonumber \\
&+& \left. \frac{(p_1^\mu\, p_2^\nu+p_1^\nu\, p_2^\mu)}{2p_3^2}\,\Delta C_{p_1p_2}+\frac{p_3^\mu\, p_3^\nu}{2p_3^4}\, \Delta C_{p_3} \right],\nonumber
\end{eqnarray}
\begin{eqnarray}
 \!\!\!\Delta C_{3g}\!\! &=&\frac {1}{4 \lambda(p_1^2,p_2^2,p_3^2)} \left [\!\! \left (\alpha_3^2-\beta_3^2\right)\log\left(\frac{\alpha_3+\beta_3}{\alpha_3-\beta_3}\right)-2 \alpha_3 \beta_3
\right],\nonumber\\
 \!\!\!\Delta C_{p_1}\!\! &=&\!\! \frac{1}{2 \lambda^2(p_1^2,p_2^2,p_3^2)}\left[(3\alpha_3^2-\beta_3^2+4 \lambda \alpha_3)\log\left(\frac{\alpha_3+\beta_3}{\alpha_3-\beta_3}\right)-2\beta_3 (3 \alpha_3 +4\lambda)\right], \nonumber\\
\!\!\!\ \Delta C_{p_3}\!\!&=&\frac{1}{4 \lambda (p_1^2,p_2^2,p_3^2)}\Big \{ \left[\beta_3^2(1-\lambda v^2)-\alpha_3^2(1-3 \lambda v^2)+2 \lambda t(t-2\alpha_3v)\right] \log\left(\frac{\alpha_3+\beta_3}{\alpha_3-\beta_3}\right ) \nonumber \\
 &+& 2\beta_3\left [\alpha_3+\lambda v(4t-3\alpha_3v) \right] \Big \},\nonumber\\ 
 \!\!\!\ \Delta C_{p_1p_3}\!\!&=&\frac{1}{2 \lambda (p_1^2,p_2^2,p_3^2)}\Big \{ \left[v(3\alpha_3^2-\beta_3^2)+2\alpha_3( \lambda v-t)-2 \lambda t\right] \log\left(\frac{\alpha_3+\beta_3}{\alpha_3-\beta_3}\right ) \nonumber \\
 &+& 2\beta_3\left [2t-v(3\alpha_3+2 \lambda) \right] \Big \},\nonumber\\ 
\!\!\!\ \Delta C_{p_1p_2}\!\!&=&-\Bigg \{ (t-\alpha_3v) \log\left(\frac{\alpha_3+\beta_3}{\alpha_3-\beta_3}\right )+\frac{2\beta_3}{\alpha_3-\beta_3}\left[t-\beta_3v+\frac{2p_3^2}{\lambda}(\alpha+4p_1^2 t) \right] \Bigg \},
\end{eqnarray} 
where 
\begin{eqnarray}
&&t=p_3^2+m_1^2-m_3^2,\nonumber\\
&&v=(p_1^2-p_2^2+p_3^2)/\lambda(p_1^2,p_2^2,p_3^2),\nonumber\\ &&\beta_3=\sqrt{\lambda(p_1^2,p_2^2,p_3^2)\lambda (p_3^2,m_1^2,m_3^2) },\nonumber\\ &&\alpha=p_1^2[p_1^2+2m_3^2-(p_2^2+p_3^2+m_1^2+m_2^2)]+(m_1^2-m_2^2)(p_3^2-p_2^2),\nonumber \\ 
&&\alpha_3= p_3^2[3p_1^2+3m_1^2+p_3^2-(p_2^2+m_3^2+2m_2^2)]+(m_1^2-m_3^2)(p_1^2-p_2^2),\nonumber\\
&&\lambda \equiv \lambda(p_1^2,p_2^2,p_3^2)\nonumber
\end{eqnarray}
 and $g^{\mu\nu}=(+,-,-,-)$ is the metric tensor. Here $m_1$ and $m_2$ represent the masses of the internal particles where we cut through. The four momenta $p_1^\mu$ and $p_2^\mu/p_3^\mu$ are the momentum of the charged Higgs and neutralino/chargino, respectively. The other type of horizontal cut can be obtained by simply replacing $p_3^\mu\leftrightarrow p_2^\mu$, and $m_1 \leftrightarrow m_2$.

 For completeness we include the terms with the vertical cut, evaluated in \cite{Frank:2007ca}
\begin{eqnarray}
\displaystyle
 \!\!\!\Delta C^\mu\!\! &=&\!\! \frac{2\pi i\, \Theta(p_1^2-(m_1+m_2)^2)}{\sqrt{\lambda(p_1^2,p_2^2,p_3^2)}}\,\left[\log\left(\frac{\alpha+\beta}{\alpha-\beta}\right)p_3^\mu
 -\left((u-\alpha v)\,\log\left(\frac{\alpha+\beta}{\alpha-\beta}\right)+2\beta v\right)\frac{p_2^\mu}{2\,p_1^2}\right],\nonumber\\
\!\!\Delta C^{\mu\nu}\!\! &=&\!\! \frac{2\pi i\, \Theta(p_1^2-(m_1+m_2)^2)}{p_1^2 \sqrt{\lambda(p_1^2,p_2^2,p_3^2)}}\,\left(\frac{g^{\mu\nu}}{2} \Delta C_g + \frac{p_2^\mu\, p_2^\nu}{8\,p_1^2}\, \Delta C_{22} + (p_2^\mu\, p_3^\nu+p_2^\nu\, p_3^\mu)\,\Delta C_{23}+p_3^\mu\, p_3^\nu\, \Delta C_{33} \right),\nonumber
\end{eqnarray}
\begin{eqnarray}
 \!\!\!\Delta C_g\!\! &=&\!\! \frac{\alpha\,\beta\, v}{4\,p_1^2}-
  (\alpha+\beta)\left(\frac{1}{2}-\frac{u}{4\,p_1^2}+\frac{(\alpha-\beta)v}{8\,p_1^2}\right)\log\left(\frac{\alpha+\beta}{\alpha-\beta}\right)
 - \beta\left(1-\frac{u}{2\,p_1^2}\right)\log\left(\frac{\alpha-\beta}{2\,p_1^2\,e}\right),\nonumber\\
 \!\!\!\Delta C_{22}\!\! &=&\!\! \left[(\alpha^2-\beta^2)\left(\frac{1}{\lambda(p_1^2,p_2^2,p_3^2)}-v^2\right)-2(u-\alpha\,v)^2\right] \log\left(\frac{\alpha+\beta}{\alpha-\beta}\right)+8\beta v(u-\alpha\,v)\nonumber\\
 &&+2\alpha\beta\left(\frac{1}{\lambda(p_1^2,p_2^2,p_3^2)}+v^2\right),\nonumber\\
\!\!\!\Delta C_{23}\!\! &=&\!\beta v + \frac{u-\alpha\,v}{2}\,\log\left(\frac{\alpha+\beta}{\alpha-\beta}\right),\nonumber\\
 \!\!\!\Delta C_{33}\!\! &=&-p_1^2\,\log\left(\frac{\alpha+\beta}{\alpha-\beta}\right)\,,
\end{eqnarray}
where
\begin{eqnarray}
u&=&p_1^2+m_1^2-m_2^2\nonumber \\
\beta&=&\sqrt{\lambda(p_1^2,p_2^2,p_3^2)\lambda (p_1^2,m_1^2,m_2^2) }\,.\nonumber\
\end{eqnarray}


\end{document}